\documentclass[useAMS,usenatbib]{mnras}
\usepackage[T1]{fontenc}
\usepackage{aecompl}
\usepackage{aas_macros}
\usepackage{pdflscape}
\newcommand{\Msol}{$M_\odot$}
\newcommand{\Rsol}{$R_\odot$}
\newcommand{\lc}{$\ell$\,Carinae}
\newcommand{\kms}{km\,s$^{-1}$}
\newcommand{\ms}{m\,s$^{-1}$}
\newcommand{\drp}{\Delta R/p}
\newcommand{\Coralie}{{\it Coralie}}
\newcommand{\Pionier}{{\it PIONIER}}

\usepackage{graphicx}
\usepackage{txfonts}
\begin{document}

\title[The Modulated Pulsations of \lc ]{Investigating Cepheid \lc 's
Cycle-to-cycle Variations via Contemporaneous
Velocimetry and Interferometry\thanks{Based on observations
collected with \Pionier\ at the ESO Very Large Telescope Interferometer located at ESO Paranal Observatory under programme number 094.D-0584 and on observations
collected with the \Coralie\ echelle spectrograph mounted to the Swiss
1.2m Euler telescope located at La Silla Observatory, Chile.} }

\author[R.I.~Anderson et al.]
{R.~I.~Anderson,$^{1,2}$\thanks{E-mail:\texttt{ria@jhu.edu}} 
A.~M\'erand,$^{3}$ 
P.~Kervella,$^{4,5}$ 
J.~Breitfelder,$^{3,4}$
J.-B.~LeBouquin,$^{6}$  \newauthor
L.~Eyer,$^{7}$
A.~Gallenne,$^{8}$ 
L.~Palaversa,$^{7}$ 
T.~Semaan,$^{7}$ 
S.~Saesen,$^{7}$ and 
N.~Mowlavi$^{7}$ 
\\\\
$^1$ Physics and Astronomy Department, The Johns Hopkins
University, 3400 N. Charles St, Baltimore, MD 21218, USA \\
$^2$ Swiss National Science Foundation Fellow \\
$^3$ European Southern Observatory, Alonso de C\'ordova 3107,
Casilla 19001, Santiago 19, Chile \\
$^4$ LESIA (UMR 8109), Observatoire de Paris, PSL, CNRS, UPMC, Univ.
Paris-Diderot, 5 pl. Jules Janssen, 92195 Meudon, France \\  
$^5$ Unidad Mixta Internacional Franco-Chilena de Astronom\'ia (UMI 3386),
CNRS/INSU, France and \\ \hspace{1mm} Departamento de Astronom\'ia,
Universidad de Chile, Camino El Observatorio 1515, Las Condes, Santiago, Chile \\   
$^6$ UJF-Grenoble 1 / CNRS-INSU, Institut de Plan\'etologie et
d'Astrophysique de Grenoble (IPAG) UMR 5274, F-38041 Grenoble, France  \\  
$^7$ Observatoire de Gen\`eve, Universit\'e de Gen\`eve, 51 Ch.
des Maillettes, 1290 Versoix, Switzerland  \\ 
$^8$ Departamento de Astronom\'ia, Universidad de Concepci\'on, 
Casilla 160-C, Concepci\'on, Chile \\ 
}

\date{Accepted 2015 October 17.  Received 2015 October 16; in original form 2015 September 21}

\pagerange{\pageref{firstpage}--\pageref{lastpage}} 

\pubyear{2015}

\maketitle 

\label{firstpage}

\maketitle

\begin{abstract}
Baade-Wesselink-type (BW) techniques enable geometric distance measurements of Cepheid variable stars in the Galaxy and the Magellanic clouds. The leading uncertainties involved concern projection factors required to translate observed radial velocities (RVs) to pulsational velocities and recently discovered modulated variability.

We carried out an unprecedented observational campaign involving long-baseline interferometry (VLTI/\Pionier ) and spectroscopy (Euler/\Coralie ) to search for modulated variability in the long-period ($P \sim 35.5$\,d) Cepheid \lc . We determine highly precise angular diameters from squared visibilities and investigate possible differences between two consecutive maximal diameters, $\Delta_{\rm{max}} \Theta$. We characterize the modulated variability along the line-of-sight using $360$ high-precision RVs.

Here we report tentative evidence for modulated angular variability and confirm cycle-to-cycle differences of \lc 's RV variability. Two successive maxima yield $\Delta_{\rm{max}} \Theta = 13.1 \pm 0.7 ({\rm stat.})\, \mu$as for uniform disk models and $22.5 \pm 1.4 ({\rm stat.})\,\mu$as ($4\%$ of the total angular variation) for limb darkened models. By comparing new RVs with 2014 RVs we show modulation to vary in strength. Barring confirmation, our results suggest the optical continuum (traced by interferometry) to be differently affected by modulation than gas motions (traced by spectroscopy). This implies a previously unknown time-dependence of projection factors, which can vary by $5\%$ between consecutive cycles of expansion and contraction.

Additional interferometric data are required to confirm modulated angular diameter variations. By understanding the origin of modulated variability and monitoring  its long-term behavior, we aim to improve the accuracy of BW distances and further the understanding of stellar pulsations.

\end{abstract}


%
%_______________________________________________________________

\section{Introduction}

Classical Cepheid variable stars are excellent probes of stellar evolution and
pulsation physics. Moreover, they are crucial calibrators of cosmic distances
and hold the key for determining the Hubble constant to within a couple of
percent accuracy \citep[e.g.][]{2011ApJ...730..119R,2012ApJ...758...24F} ,
aiding in the interpretation of the cosmic microwave background for cosmology. 
\citep{2013ApJS..208...19H,2014A&A...571A..16P} 

Large efforts are currently under way to push the accuracy of a measurement of
the local Hubble constant, $H_0$, to $1\%$
\citep{2012arXiv1202.4459S}. Improved Cepheid distances are
a crucial element in pursuit of such extreme cosmological precision. Extragalactic
Cepheid distances are typically determined using period-luminosity relations
\citep[PLRs,][]{1908AnHar..60...87L,1912HarCi.173....1L} whose calibration
is achieved more locally, e.g. in the Galaxy
\citep{1997MNRAS.286L...1F,2007MNRAS.379..723V,2007AJ....133.1810B}, the Large
Magellanic Cloud
\cite[e.g.][]{2008AcA....58..163S,2013A&A...550A..70G,2015AJ....149..117M}, the
mega-maser galaxy NGC\,4258
\citep{2006ApJ...652.1133M,2013ApJ...775...13H,2015AJ....149..183H}, or
combinations of these \citep[e.g.][]{2011ApJ...730..119R}. One remaining key
issue in this context is how chemical composition affects the calibration of
PLRs. 

An investigation into the impact of metallicity on slope and zero-point of PLRs
requires distance measurements to individual Cepheids of different metallicities.
While ESA's space mission {\it Gaia} will soon provide an unprecedented data set
of extremely accurate Cepheid parallaxes, the metallicity
range spanned by this sample will be relatively small. Extending the sample to
Cepheids in the Magellanic clouds would provide a stronger handle on
metallicity. However, even {\it Gaia} will not be able to measure individual
Cepheid parallaxes at such great distances with the accuracy required to
separate depth effects inside the LMC from metallicity effects on PLRs.
This point is particularly difficult for the Small Magellanic Cloud \citep{2015arXiv150206995S},
which provides the greatest lever for metallicity. 

Fortunately, variants of the Baade-(Becker-)Wesselink
\citep[BW;][]{1918MNRAS..78..639L,1926AN....228..359B,1940ZA.....19..289B,1946BAN....10...91W}
technique afford a homogeneous methodology for determining individual geometric
distances to Cepheids in the Galaxy, LMC, and SMC
\citep[e.g.][]{2004A&A...415..531S,2008A&A...488...25G,2011A&A...534A..94S,2013A&A...550A..70G}.
However, BW distances have suffered from systematic difficulties related to the
calibration of the projection factor as described below. This has limited their
ability to reveal how metallicity affects the PLR.
It is therefore important to improve the accuracy of BW distances, and new
observational opportunities such as long-baseline near-infrared (NIR) interferometry
\citep[e.g.][]{1997A&A...320..799F,2000ApJ...543..972N,2001A&A...367..876K,2004A&A...416..941K,2004A&A...423..327K}
and infrared surface brightness relations \citep{2004A&A...428..587K} are
providing ever higher precision.

BW techniques exploit pulsational motion to measure {\it geometric} distances.
In essence, distance
\begin{equation}
d\ \rm{[pc]} = 9.3095 \cdot \Delta R [R_\odot] / \Delta \Theta \rm{[mas]}\ ,
\label{eq:distance}
\end{equation}
where $\Delta R$ denotes the {\it linear} radius variation in \Rsol\
\citep[using \Rsol=$696,342 \pm 65$\ km from][]{2012ApJ...750..135E} as measured
from radial velocities (RVs) and $\Delta \Theta$ the angular diameter variation
in {\it milli}arcseconds. The method's accuracy rests to a large extent on the
empirically calibrated projection
factor $p$ (see Breitfelder et al., in preparation), since $p$ is required to
translate observed line-of-sight velocities (which integrate over a limb-darkened disk) to the pulsational velocity as seen
from the star's center. The linear equivalent to $\Delta \Theta$ is thus
obtained by computing
\begin{equation}
\Delta R = p \cdot \int_{t_1}^{t_2} v_r \rm{d}t \ .
\label{eq:deltaRfull}
\end{equation}

\citet{2007A&A...471..661N} decomposed this projection factor as 

\begin{equation}
p = p_o f_{\rm{grad}} f_{\rm{o-g}}\ , 
\label{eq:pfactor}
\end{equation} 
where geometric aspects of limb darkening and disk integration
are included in $p_o$, velocity differences between the {\it optical} and {\it
gas} photospheric layers are represented by $f_{\rm{o-g}}$, and 
velocity gradients acting over the line forming region by $f_{\rm{grad}}$.

\citet{1995ApJ...446..250S} and \citet{2004A&A...428..131N} investigated
how the characteristic, variable spectral line asymmetry of Cepheids introduces
a (repeating) phase-dependence of projection factors. It is generally assumed,
however, that these variations of $p$ cancel out when integrating over an entire
pulsation cycle.

Recently, \citet{2014A&A...566L..10A} demonstrated an additional difficulty for
determining BW distances. Four Cepheids were shown to exhibit modulated RV
curves, where modulation refers to differences in RV curve shape that occur on
time-scales from weeks to years. RV curve modulation can lead to significant
differences in
\begin{equation}
\Delta R / p = \int_{t_1}^{t_2} v_r \rm{d}t
\label{eq:deltaR}
\end{equation}
determined from consecutive pulsation cycles for long period Cepheids;
short-period (likely overtone) Cepheids exhibit modulation on longer time-scales.
Specifically, \lc\ exhibited a systematic difference of $5-6\%$ between
consecutive pulsation cycles.
We stress that this effect is notably different from the
\emph{phase}-dependent $p$-factors discussed by \citet{1995ApJ...446..250S}.
\citet{2014A&A...566L..10A} discussed RV curve modulation in terms of a
systematic uncertainty for BW distances, since $\drp$ linearly enters $d$ in
Eq.\,\ref{eq:distance}. It remained unclear, however, how $\Delta \Theta$
relates to cycle-to-cycle changes in $\drp$. 

To answer this question, we started a 
monitoring campaign involving contemporaneous long-baseline interferometric and
spectroscopic observations that operated between 2014 December 20 and 2015 May 22. \lc\ was
chosen as the target for this campaign, since it subtends the largest angular
size of any known Cepheid on the sky and since previous observations by
\citet{2004ApJ...604L.113K} and \citet{2009MNRAS.394.1620D} indicated that the
instrumental precision of modern optical/NIR interferometers may be sufficient to reveal cycle-to-cycle
differences at the order or magnitude indicated by the RV modulations.

In this paper, we investigate whether cycle-to-cycle differences are present in
both the {\it angular} motion of {\it optical} layers traced by interferometry,
and the {\it line-of-sight} or linear motion of {\it gas} layers traced by RVs,
cf. Eq.\,\ref{eq:pfactor}. We describe our new observations in
Sect.\,\ref{sec:obs}. Using these new observations, we first determine angular
diameters from squared visibility curves in Sect.\,\ref{sec:interferometry}. We
then study the RV variability in Sect.\,\ref{sec:spec} 
using RVs determined by cross-correlation.
We discuss our findings and their implications in Sect.\,\ref{sec:discuss} and
conclude in Sect.\,\ref{sec:conclude}. We provide supporting
materials in Appendix A available in the electronic version of the journal.

%__________________________________________________________________

\section{Observations}\label{sec:obs}

We have gathered an unprecedented data set comprising $360$
high-precision RVs derived from optical spectra and $15$ 
high-precision angular diameters from NIR long-baseline
interferometry. In this section, we describe the observations and resulting
data sets.

\subsection{VLTI/\Pionier\ High-precision Long-baseline Near-IR
Interferometry}
\label{sec:obs:pionier} 

We carried out interferometric observations with the four-beam combiner
\Pionier\ \citep{2011A&A...535A..67L} at the Very Large Telescope Interferometer
\citep[VLTI, cf.][]{2014SPIE.9146E..0JM}
during three observing runs in 2015 January and February (ESO Program ID
094.D-0583).
The observations were taken near three epochs of predicted extremal diameter comprising two maxima and one minimum. We refer to these as max1,
min1, and max2 in the remainder of this paper.

Our observations were aimed at determining angular diameters 
at successive extrema to test whether consecutive extrema would be significantly
different, and how the full amplitude of angular diameter
variations would behave if closely monitored.
The motivation to carry out this investigation was given by observed cycle-to-cycle differences in the RV variability of four Cepheids
\citep{2014A&A...566L..10A}.

We carried out all observations using all four $1.8$\,m Auxiliary Telescopes
(ATs) placed to achieve the longest available baselines (up to $140$\,m,
referred to here in units of mega-$\lambda$),
aiming for the highest spatial frequencies to enable maximal spatial
resolution.
This was crucial for achieving the precision required for this program, since the longest baselines are just about capable of
resolving the first zero of the visibility curves of \lc\ near its maximal
diameter.

All measurements were taken with the recently installed {\it RAPID}
detector\footnote{\url{http://www.eso.org/public/usa/announcements/ann15042/}},
which is optimized for fainter targets. In the medium sensitivity setting, we
employed typical read-out times of 0.5\,ms, which enabled up to $100$ science
object observations per half-night.  All observations were taken in the new
{\tt GRISM} dispersion mode, which supplies six spectral channels in the {\it H} band
($1.53$, $1.58$, $1.63$, $1.68$, $1.73$, and $1.78\,\mu$m) with a spectral
resolving power $R \sim 45$. Our typical (nightly) visibility curves for \lc\
contain approximately $500$ data points.

We applied this extreme strategy in order to achieve the highest possible
sensitivity to even the most minute cycle-to-cycle differences in the angular
diameters.
We followed the well-established observing strategy of alternating between
observations of the science target and calibrator stars. During the first
observing run (run A), this sequence was {\tt Cal1, Sci, Cal2, Sci, Cal1, etc.}
During later observing runs, we added additional calibrator stars. The
calibrators common to all observations are HD\,81502 and HD\,89805. We calibrate our
science observations using only the calibrator stars in common to all three
observing runs, namely HD\,81502 and HD\,89805, and employ two other
calibrators, namely HD\,74088 and HD\,81011, as \emph{standard stars}, see the online Appendix\,\ref{app:A}. Standard star observations are crucial for 
demonstrating that the precision and instrumental stability achieved are
sufficient for identifying apparent cycle-to-cycle differences as features of
the pulsations.

Table\,\ref{tab:cals} lists all calibrator stars observed during our program.
These were selected using the {\tt SearchCal} tool
\citep{2006A&A...456..789B,2011A&A...535A..53B}
and/or taken from the catalog by \citet{2005A&A...433.1155M} based on proximity
on the sky, similar magnitude, and (where possible) accurately known
diameters. Calibrators HD\,90853, HD\,102461, HD\,102964, and
HD\,110458 are not used in this work, as they were observed only
sporadically and/or variable stars.

\begin{table}
\centering
\begin{tabular}{@{}lrrlrrl@{}}
\hline
HD & Sep & $m_H$ & SpTyp & $\Theta_{\rm{UD}}$ & $\sigma(\Theta_{\rm{UD}})$ &
Run
\\
 & [deg] & [mag] & & [mas] & [mas] \\
\hline
74088$^\dagger$ & 7.65 & 3.15 & K4III & 1.493$^a$ & 0.019 & C\\  
81101$^\dagger$ & 2.81 & 2.67 & G6III & 1.394$^b$ & 0.099 & B,C\\
\textbf{81502} & 3.42 & 3.24 & K1.5II-III & 1.230$^a$ & 0.016 & A,B,C\\
\textbf{89805} & 4.61 & 2.87 & K2II & 1.449$^a$ & 0.019 & A,B,C\\
90853$^1$ & 6.46 & 2.96 & F2II & 1.025$^b$ & 0.072 & B\\
102461$^2$ & 16.2 & 1.66 & K5III & 2.93$^c$ & 0.034 & B\\
102964$^\ddagger$ & 26.7 & 1.69 & K3III & 2.51$^c$ & 0.039 & B\\
110458$^\ddagger$ & 30.1 & 2.38 & K0III & 1.65$^c$ & 0.018 & B\\
\hline
\end{tabular}
\caption{Stars of known angular diameter observed contemporaneously with \lc .
The two stars listed in bold font, HD\,81502 and HD\,89805, were observed during each observing
run and are used to calibrate the measurements of \lc\ and any standard stars
identified by $^\dagger$. Calibrators HD\,102964 and HD\,110458 (marked with
$^\ddagger$) were disregarded due to their small number of observations.
Notes on individual stars. $^1$: {\it s}\,Carinae (not to be confused with variable star S\,Carinae) is a possible low-amplitude variable star \citep{1959MNSSA..18...48S} of unknown type.
$^2$: V918\,Centauri, a variable star of unknown type with ({\it
Hipparcos}-band) amplitude $0.0216$\,mag and frequency $f=0.05517$
\citep{2002MNRAS.331...45K}. 
Origin of literature diameters: $^a$ \citet{2005A&A...433.1155M}, $^b$
\citet{2010SPIE.7734E..4EL}, $^c$ \citet{2002A&A...393..183B}. }
\label{tab:cals}
\end{table} 

The data reduction and calibration was performed using the \Pionier\ pipeline
{\tt pndrs}, adapted for the new detector.
We cleaned the nightly time series, rejecting data that were obviously affected
by meteorological conditions or technical problems such as the failing delay
line DL4 (night of 2015 January 12) or if fringe tracking was temporarily lost.
We removed a total of $700$ of $10238$ UV points\footnote{Each observation
supplies six UV points}.
Once we completed outlier rejection, we used the \Pionier\ pipeline to compute
squared visibilities $V^2$, triple product amplitudes, and triple product
closure phases. In this work, we focus mainly on the derived squared
visibilities. We studied \lc 's closure phase data using the {\tt python} tool
{\tt CANDID}\footnote{\url{https://github.com/amerand/CANDID}}
\citep{2015A&A...579A..68G} to search for a close companion or signs of
asymmetry, cf. Sect.\,\ref{sec:disc:companion}. We also inspected closure phase
stability of the calibrators and standard stars to ensure that these objects did
not show any signs of companions.

\begin{landscape}
\begin{figure}
\includegraphics{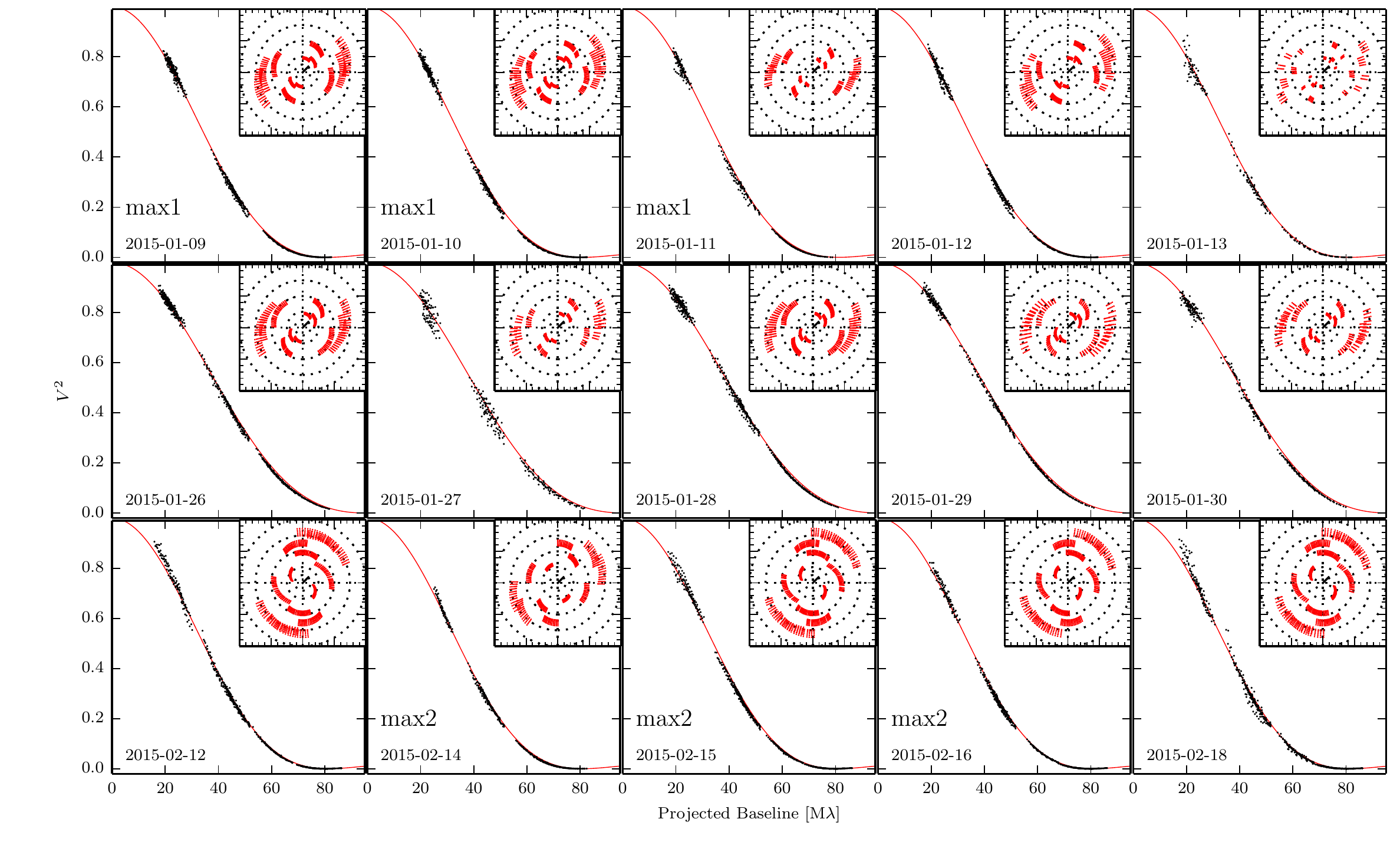}
\caption{Squared visibilities against projected baselines in mega-$\lambda$ 
and UV plane coverage for all nights of program 094.D-0583. The axis range
is $0 - 95$\,M$\lambda$ on the $x$-axis and $-0.02$ to unity on the $y$-axis. The local
date of the beginning of each night is shown in the bottom left. The top panels
show the epoch near the first maximum, followed by the epoch
near minimal radius in the center panels, and the second maximum in the bottom
row. The best-fitting uniform disk diameter fit is shown as a solid red curve. Nights
near the first and second maximum are marked as max1 and max2, respectively. The
UV-plane polar circles indicate 12.5, 25, 50, 75, and 100 M$\lambda$.}
\label{fig:vlti:visibilities}
\end{figure}
\end{landscape}

Figure\,\ref{fig:vlti:visibilities} shows the calibrated squared visibilities
against projected baseline for the measurements of this program. Each row
represents one epoch near an extremal diameter (max1, min1, max2) and the
UV-plane coverage is shown in the top right corner of each night.
Fig.\,\ref{fig:vlti:visibilities} demonstrates the exquisite data quality
achieved and helps to identify nights of better and worse quality. Uninterrupted
observations and densely-covered UV-planes identify the better nights. 
The most crucial nights for comparing maximal radii were the three nights from
2015-01-09 through 2015-01-11 (marked max1) and 2015-02-14 through 2015-02-16
(marked max2). 

Figure\,\ref{fig:vlti:visibilities} also shows that the baseline configurations
used are sufficient to resolve the first zero of the squared visibility curve
during epochs near maximal diameter. Near minimal diameter, however, the first
zero is not fully resolved.

We were extremely lucky not to suffer any completely lost night over the
entirety of our 15 night observing program, with a total of $5$ hours lost due
to weather.
Nevertheless, our data are of course subject to ambient conditions and other
circumstances such as technical issues.
We also note the failure of delay line 4 during observations on the night of 12
January. Inspection of the $V^2$ curve from that night and the resulting
diameter indicates that this night's measurements are offset (biased) with
respect to the other nights. While we cannot trace the reason for this
outlier, we note a peculiar bifurcation in the visibility curve
(Fig.\,\ref{fig:vlti:visibilities}) and the systematically higher $\chi^2$
values (compared to the other nights of this run) returned by the diameter
fitting procedures (cf. Sect.\,\ref{sec:interferometry}).
This does not constitute a problem for the present work, however, since we
investigate primarily the three nights closest to maximum, see below.
Table\,\ref{tab:vlti:quality} provides an observing log for each night.

\begin{table}
\centering
\begin{tabular}{@{}llrrrrr@{}}
\hline
Night & BL & Sky & Seeing & $\tau_{\rm{coh}}$ & $t_{\rm{l,W}}$ & Tech \\
 & & &  [arcsec] & [ms] & [h] & \\
\hline
9 Jan & A1-G1-K0-I1 & TN &  $0.7 - 1$ & $1 - 2$ &  & \\
10 Jan & A1-G1-K0-I1 & TK &  $1 - 1.5$ & $\sim 1$ & &  \\
11 Jan & A1-G1-K0-I1 & CY-TN & $0.8 - 1.5$ & $2 - 1$ & 3 &  \\
12 Jan & A1-G1-K0-I1 & TN &  $0.7 - 1.5$ & $< 2$ &  & DL4 \\
13 Jan & A1-G1-K0-I1 & TK & $> 1$ & $\sim 1$ & 2 & \\
\hline
26 Jan & A1-G1-K0-I1 & CL-WI & $0.5-0.8$ & $4-8$ &  &   \\
27 Jan & A1-G1-K0-I1 & CL-PH & $1.5 - 2.5$ & $\sim 1$ &  &  \\
28 Jan & A1-G1-K0-I1 & CL &  $0.5 - 1.2$ & $2 - 3$ & &  \\
29 Jan & A1-G1-K0-I1 & TN-CL & $1 - 1.5$ & $<2$ & &    \\
30 Jan & A1-G1-K0-I1 & TK-CL & $0.5 - 1.3$ & $\sim 2$ & &   \\
\hline
12 Feb & A1-G1-K0-J3 & PH & $\sim 1$ & $\sim 2$ &  &   \\
14 Feb & A1-G1-K0-J3 & PH & $1.7 - 1$ & $2 - 4$ &  &   \\
15 Feb & A1-G1-K0-J3 & PH & $0.5 - 1$ & $4.5 - 3$ &  &  \\
16 Feb & A1-G1-K0-J3 & PH & $1.3 - 1$ & $2 - 4$ &  &   \\
18 Feb & A1-G1-K0-J3 & HU CL & $1 - 2$ & $< 1$ &  &   \\ \hline
\end{tabular}
\caption{Observation log for our VLTI/\Pionier\ program. Column BL lists
the baseline configuration of the ATs, see
\url{http://www.eso.org/sci/facilities/paranal/telescopes/vlti/configuration/P94.html.html}
for more details. The information given for sky quality, seeing, etc. is given
in chronological order per night. Sky quality is abbreviated as follows:
TN = thin clouds, TK = thick clouds, CL = clear, CY = cloudy, WI = strong winds,
PH = photometric, HU = high humidity. Column Seeing gives the  range of DIMM
seeing measured over the course of our observations. 
$\tau_{\rm{coh}}$ is the coherence time-scale of the atmosphere, $t_{\rm{l,W}}$
lists the time lost due to weather in hours, and Column Tech indicates nights
with technical difficulties. Additional details are available through the ESO
archive at \url{http://archive.eso.org/asm/ambient-server}. }
\label{tab:vlti:quality}
\end{table}

\subsection{\textit{Coralie} High-resolution Optical Spectra}
\label{sec:obs:coralie}

We conducted optical high-resolution spectroscopic observations of \lc\ using
the \Coralie\ spectrograph \citep{2001Msngr.105....1Q} at the Swiss $1.2$\,m Euler telescope located at La Silla Observatory, Chile. These observations served three primary purposes: (a) to quantify the line-of-sight component of \lc 's pulsations, i.e., $\drp$ (cf. Eq.\,\ref{eq:deltaR}), contemporaneously with the interferometric measurements; (b) to confirm the cycle-to-cycle modulations seen in 2014; (c) to provide precise timings for each cycle of contraction, expansion or full pulsation cycle (contraction and expansion).
We typically aimed for three measurements per night, spaced out over the
visibility of the object from La Silla Observatory to limit phase gaps
introduced by the day/night cycle.
Steeper and extremal parts of the RV curve were monitored more closely, with up
to 10 observations in a given night. Our \Coralie\ monitoring campaign began on
20 December 2014 and ended on 22 May 2015.

The raw data were reduced using the \Coralie\ pipeline, which includes pre- and
over-scan bias corrections, flat-fielding using halogen lamps, cosmic-clipping,
and order extraction. The wavelength calibration is provided by thorium-argon
lamp reference spectra. All observations were made in the OBTH observing mode,
wherein a Th-Ar spectrum is interlaced between the orders of the science target
via a secondary fiber to provide simultaneous corrections for RV
zero-point drifts. This observing mode offers the highest precision. We
determine RVs via the cross-correlation technique
\citep{1996A&AS..119..373B,2003ASPC..294...39P} using a numerical line mask
representative of a solar spectral type. Table\,\ref{tab:RVs:data} shows a subset of our new RV measurements of $\ell$\,Carinae. The full data set is available in the electronic version of the journal as well as via the CDS.

\begin{table}
\centering
\begin{tabular}{lrr}
\hline
BJD  $- 2\,400\,000$ & $v_r$ & $\sigma (v_r)$ \\
d & [\kms ]  & [\kms ] \\ 
\hline
57012.630987 & 19.603 & 0.005  \\
57012.688859 & 19.623 & 0.004  \\
57012.754012 & 19.636 & 0.004  \\
57012.790623 & 19.630 & 0.004  \\
57012.866436 & 19.628 & 0.004  \\
57013.686595 & 19.259 & 0.004  \\
57013.736319 & 19.221 & 0.004  \\
57013.787895 & 19.174 & 0.004  \\
57013.834101 & 19.106 & 0.003  \\
57013.865421 & 19.069 & 0.004  \\
\multicolumn{3}{c}{\ldots} \\
57158.496217 &  4.944 & 0.004  \\
57158.641075 &  3.316 & 0.004  \\
57159.494445 & -4.971 & 0.004  \\
57159.628088 & -5.962 & 0.004  \\
57160.493287 & -10.866 & 0.004  \\
57160.668815 & -11.552 & 0.004  \\
57162.489324 & -15.003 & 0.004  \\
57162.629240 & -15.042 & 0.004  \\
57164.525810 & -14.314 & 0.004  \\
57164.623180 & -14.228 & 0.012  \\
\hline
\end{tabular}
\caption{Example data from our new (2014-2015) \Coralie\ RV campaign.
The first and last 10 measurements are shown. The full data set is made publicly
available in the electronic version of the journal and through the
CDS.}
\label{tab:RVs:data}
\end{table}

\Coralie\ received an upgrade in November 2014, before our
monitoring began. During this upgrade, octagonal fibers were installed to
replace the previously installed fibers with circular cross-sections, and a
double-scrambler was reintroduced (F. Pepe 2014, private communication).
Octagonal fibers have been shown to yield higher precision RVs than circular
ones \citep{2013A&A...549A..49B}, which is primarily attributed to better light
scrambling as the light passes through the fiber into the spectrograph.
The realignment of the optical path and the new optical fiber are expected to
introduce a small zero-point offset with respect to \Coralie\ RVs measured
prior to the upgrade. Based on a monitoring of RV standard stars, this offset is
on the order of $15$\,\ms, but may depend on spectral type (F. Pepe private
communication), and thus, may vary slightly as a function of pulsation phase.
Since the upgrade was carried out before the RV monitoring campaign began, we expect no significant zero-point
differences among the observations taken for this program. There may, however,
exist small ($\sim 15\,$\ms), possibly phase-dependent, differences between the
\Coralie\ RVs published by \citet{2014A&A...566L..10A} and the present ones.

RV uncertainties are determined by the \Coralie\ pipeline and take into account
photon noise as well as the shape of the CCF. In the case of the very bright
\citep[$\langle m_V \rangle \sim 3.75\,$mag,][]{1995IBVS.4148....1F} star \lc, 
the derived uncertainty is limited by photon noise \citep{2001A&A...374..733B}.
The resulting typical uncertainties are on the order of a few \ms, with 
median $3.14$\,\ms. This is five times better than
the measurements published in \citet{2014A&A...566L..10A}, owing primarily to
the different observing mode (OBTH).

We note that Cepheid atmospheres are much more complex than the atmospheres
of stars to which this level of instrumental precision is usually applied, being
subject to velocity gradients, turbulence, large scale convection,
granulation, and possibly shock. In light of this, the above-stated
uncertainties should not be mistaken for estimates of {\it accuracy}
\citep[cf.][]{2003A&A...401.1185L}, both because the measurement is biased
and because no unique velocity applies to the entire atmosphere of the star.

In this work, we aim to determine differences in the behavior of pulsation
cycles, i.e., we aim for the utmost {\it precision}. To this end, we adopt the
definition that leads to the most {\it precise} estimate of (a weighted
atmospheric average) RV. As shown by \citet{2013PhDT.......363A}
and reflected by the low scatter (during a single pulsation cycle) of the RV
data presented here, so-called Gaussian RVs determined with the
cross-correlation technique yield the most precise estimation of such mean
velocities. We therefore adopt -- as is also common practice -- Gaussian RVs in
this work. Further work (Anderson, in preparation) will provide additional
insight into these questions.

\section{Results}
We report the results of our observational program in this section, starting
with the interferometric part in Sect.\,\ref{sec:interferometry}, followed by
the results from spectroscopy in Sect.\,\ref{sec:spec}. 

\subsection{Modulated Interferometric Variability}
\label{sec:interferometry}
\label{sec:AngDiam}
The primary purpose of our observations for this paper is to obtain a
differential measurement of \lc 's diameter determined near two 
consecutive maxima. To achieve the greatest possible precision, we obtained
an unprecedented amount of observations with a state-of-the-art instrument, and 
aimed at avoiding observational biases as much as possible, e.g.
by using the same calibrator stars for all observations.

We determined angular diameters $\Theta$ by fitting squared visibility
curves assuming both uniform disk ($\Theta_{\rm{UD}}$) and limb-darkened models.
We determine $\Theta_{\rm{UD}}$ as well as diameters assuming linear limb
darkening $\Theta_{\rm{LD,lin}}$ using {\tt LITpro}
\citep{2008SPIE.7013E..1JT}, a tool made freely available by the {\tt
JMMC}\footnote{\url{http://www.jmmc.fr/litpro_page.htm}}.
We adopt the fixed coefficient in the $H$ band $u_{\rm{LD,H}} = 0.29$ for models
assuming linear limb darkening \citep{2013A&A...554A..98N}. 

We also determine limb darkened (Rosseland) diameters,
$\Theta_{\rm{SATLAS}}$, based on spherical {\tt SATLAS} models
\citep{2013A&A...554A..98N}, adopting a temperature range of $4700 - 5300\,$K
for the phases covered by our observations
\citep{2011AJ....142...51L}. To this end, we compute $V^2$ profiles using a
numerical Hankel transform, taking into account bandwidth smearing effects.
These limb-darkened diameters are likely the most {\it accurate} ever
determined for a Cepheid variable star, given the amount of high-quality data
used, mainly limited by assumptions on limb darkening and the wavelength solution.
While we do adopt \emph{phase-dependent} limb darkening laws (by adopting a
range of $T_{\rm{eff}}$), we do not assume a \emph{cycle-dependent} limb
darkening law, since cycle-to-cycle differences in temperature at fixed
pulsation phase are at least an order of magnitude smaller and will not
significantly impact the result.
 
Table\,\ref{tab:vlti:diams} lists the time series of the diameters we
determined, as well as two measurements taken in between our program's runs,
when \lc\ was observed as a backup target\footnote{We thank F. Anthonioz for
observing \lc\ as a backup target during ESO programme ID 094.C-0884(A).}
The uncertainties listed in Tab.\,\ref{tab:vlti:diams} are formal statistical
uncertainties, i.e., do not take into account systematics such as the
uncertainty of the wavelength calibration. 

\begin{figure*}
\centering
\includegraphics{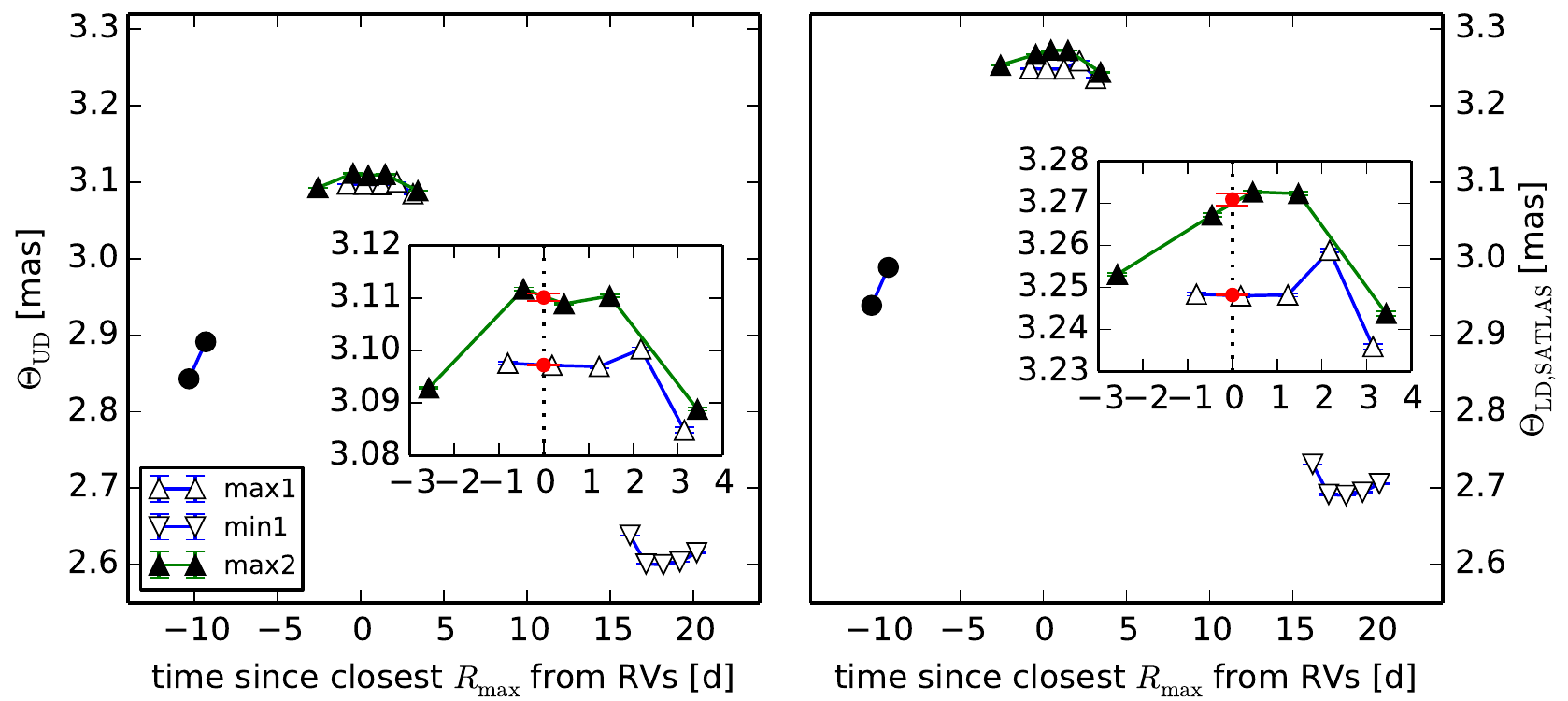}
\caption{Tentative evidence for modulated nature of \lc 's variable angular
diameters. We show diameters against time since the closest maximal {\it linear} 
radius as measured from the RV curve. Measurements near minimal,
average, and maximal diameters are plotted as downward triangles, circles, and
upward triangles, respectively. Open symbols denote earlier measurements. We
show uniform disk diameters $\Theta_{\rm{UD}}$ in the left-hand panel and
limb-darkened diameters (based on SATLAS models, cf.
Tab.\,\ref{tab:vlti:diams}) in the right-hand panel.
The measurements near maximum diameter suggest a systematic difference
between the two consecutive observed maxima.
The inlays in the center right of each panel are close-ups near maximum
$\Theta$. The mean of the three measurements near each maximum is overplotted
at time 0 as a red filled circle with the standard mean uncertainty determined from the dispersion of the points. The upward-offset open triangle near time $2$ should be given low weight due to technical problems during that night, cf.
Tabs.\,\ref{tab:vlti:quality} and \ref{tab:vlti:diams}.} 
\label{fig:vlti:maxima}
\end{figure*}

Since observations taken during the three nights closest to the maximal
diameter do not vary significantly, we average these to determine more precise
mean maximal diameters $\langle \Theta_{\rm{max1}} \rangle$ and $\langle \Theta_{\rm{max2}} \rangle$, cf. Tab.\,\ref{tab:vlti:meandiams}. We 
determine the standard mean error from the dispersion around that value, divided
by $\sqrt{3}$. Figure\,\ref{fig:vlti:maxima} shows the angular diameters from
Tab.\,\ref{tab:vlti:diams} versus time relative to the closest time of maximal
{\it linear} radius as determined from the RV curve, see
Sect.\,\ref{sec:spec:RVs}. The enlarged inlays show the clearly offset, larger
diameters of the second epoch near maximal diameter.

We thus determine a larger mean maximal diameter for the second epoch, 
differing by approximately $22.5 \pm 1.4 (\rm{stat.})\,\mu$as, or $0.7\%$ of
the second epoch's diameter. We note that the
cycle-to-cycle difference in maximal diameter is also apparent for UD
diameters and diameters determined using a linear limb darkening law.
While this difference is much larger than the squared sum of the scatter, we
caution that other effects such as ablation due to rotation, small companions,
or lack of instrumental stability could in principle introduce effects of this
order. While we carried out several investigations into the robustness of
this result, we unfortunately cannot \emph{demonstrate} the long-term stability
due to a lack of standard star observations during run A. We therefore
conservatively consider this result \emph{tentative} evidence for modulated
angular variability.

\begin{table*}
\centering
\begin{tabular}{@{}lrrrrrrrrr@{}}
\hline
MJD & $\Theta_{\rm{UD}}$ & $\sigma(\Theta_{\rm{UD}})$ & $\chi^2_{\rm{r,UD}}$ &
$\Theta_{\rm{LDlin}}$ & $\sigma(\Theta_{\rm{LDlin}})$ & $\chi^2_{\rm{r,LDlin}}$
& $\Theta_{\rm{SATLAS}}$ & $\sigma(\Theta_{\rm{SATLAS}})$ &
$\chi^2_{\rm{r,SATLAS}}$
\\
   & [mas] & [mas] & & [mas] & [mas] & & [mas] & [mas] &  \\
\hline
57032.304901 & 3.09757 & 0.00029 & 2.13 & 3.1878 & 0.00037 & 3.10 &
3.2484$^\dagger$ & 0.0004 & 4.153 \\
57033.295435 & 3.09717 & 0.00026 & 1.48 & 3.1873 & 0.00032 & 2.12  &
3.2480$^\dagger$ & 0.0004 & 2.930 \\
57034.347907 & 3.09695 & 0.00031 & 1.49 & 3.1872 & 0.00036 & 1.85 &
3.2482$^\dagger$ & 0.0004 & 2.606 \\
{\it 57035.282871} & {\it 3.10028} & {\it 0.00036} & {\it 2.89} & {\it 3.1948} &
{\it 0.00044} & {\it 3.89} & {\it 3.2588$^\dagger$} & {\it 0.0005} & {\it 5.589}
\\
57036.252418 & 3.08485 & 0.00052 & 1.33 & 3.1751 & 0.00060 & 1.62  &
3.2359$^\dagger$ & 0.0007 & 2.089 \\
\hline
57049.298269 & 2.63828 & 0.00028 & 0.82 & 2.7076 & 0.00030 & 0.91  &
2.7313$^\ddagger$ & 0.0003 & 0.963 \\
57050.260437 & 2.60085 & 0.00092 & 2.03 & 2.6687 & 0.00095 & 2.02   &
2.6918$^\ddagger$ &0.001 & 2.019\\
57051.304289 & 2.59989 & 0.00036 & 1.29 & 2.6675 & 0.00039 & 1.43   &
2.6905$^\ddagger$ &0.0004 &1.498\\
57052.301669 & 2.60375 & 0.00026 & 0.76 & 2.6716 & 0.00027 & 0.81   &
2.6947$^\ddagger$ &0.0003 &0.844\\
57053.308108 & 2.61558 & 0.00038 & 1.10 & 2.6829 & 0.00040 & 1.15  &
2.7058$^\ddagger$ & 0.0004&1.183 \\
\hline
57058.348365 & 2.84296 & 0.00247 & 0.91 & 2.9106 & 0.00255 & 0.92  & 2.9390$^*$
& 0.0026 & 0.923 \\
57059.356773 & 2.89133 & 0.00353 & 1.14 & 2.9599 & 0.00363 & 1.14   & 2.9887$^*$
&0.0037 &1.137\\
\hline
57066.101811 & 3.09287 & 0.00021 & 2.66 & 3.1883 & 0.00026 & 3.74   &
3.2531$^\dagger$ & 0.0003 & 5.128 \\
57068.216931 & 3.11169 & 0.00030 & 3.10 & 3.2044 & 0.00034 & 3.76   &
3.2672$^\dagger$ & 0.0004 & 4.854 \\
57069.127975 & 3.10887 & 0.00019 & 3.01 & 3.2064 & 0.00025 & 4.35   &
3.2726$^\dagger$ & 0.0003 & 6.186 \\
57070.147921 & 3.11036 & 0.00027 & 3.62 & 3.2067 & 0.00034 & 4.85   &
3.2723$^\dagger$ & 0.0004 & 6.55 \\
57072.109906 & 3.08883 & 0.00031 & 1.54 & 3.1813 & 0.00041 & 2.44   &
3.2438$^\dagger$ & 0.0005 & 3.418 \\
\hline
\end{tabular}
\caption{Angular diameters determined from our \Pionier\ observations.
MJD is modified Julian date, i.e., JD $-2\,400\,000.5$. Subscripts UD and LD
indicate uniform disk and limb darkened disk models, uncertainties
given are the (underestimated, see the text in Sect.\,\ref{sec:AngDiam}) statistical 
uncertainties, and all $\chi^2$ values have been divided by the
number of degrees of freedom. The measurements are grouped according to the
epochs (max1, min1, in between, max2). Values shown in italics should be
considered unreliable due to a technical problem with delay line 4. We
adopt the following $T_{\rm{eff}}$ for the SATLAS models
\citep{2013A&A...554A..98N}: $^\dagger$:
$4700$\,K, $^*$: $5100$\,K, and $\ddagger$: $5300$\,K \citep{2011AJ....142...51L}.}
\label{tab:vlti:diams}
\end{table*}

\begin{table}
\centering
\begin{tabular}{@{}lrrrr@{}}
\hline
Epoch & $\langle \Theta_{\rm{UD}} \rangle$ &
$\sigma(\langle\Theta_{\rm{UD}}\rangle)$ &
 $\langle \Theta_{\rm{SATLAS}} \rangle$ &
$\sigma(\langle\Theta_{\rm{SATLAS}}\rangle)$ \\
 & [mas] & [mas] & [mas] & [mas] \\
\hline
max1 & 3.0972 & 0.0001 & 3.2482 & 0.0001 \\
max2 & 3.1103 & 0.0007 & 3.2707 & 0.0014 \\
\hline
$\Delta_{\rm{max}} \Theta$ &  $0.0131$ & $0.0007$ & $0.0225$ & $0.0014$ \\
\hline
\end{tabular}
\caption{Mean UD and limb-darkened diameters (based on SATLAS
models, cf. Tab.\,\ref{tab:vlti:diams}) determined from the three nights closest
to the maximal epoch together with their respective standard mean errors.
The bottom row lists the difference in maximal diameter (max2 $-$ max1) and
the square-summed standard mean errors. The standard mean uncertainties do
not reflect possible additional systematic uncertainty, see text. The difference
in maximal limb-darkened diameters translates to a cycle-to-cycle difference of
$1.2\,$\Rsol .} 
\label{tab:vlti:meandiams}
\end{table}

At the extreme level of precision aimed for in this work, all possible sources
of bias and instrumental effects must be taken into account
\citep[e.g.][]{2004SPIE.5491..741K}. These include non-linear and time-dependent effects that can introduce unknown, albeit possibly significant, bias. Both types of systematics can in principle be traced
by observations of standard stars. Time-dependent effects include shorter
(intra-run) and longer (inter-run) effects. We performed the following tests to
investigate whether the cycle-to-cycle differences are likely to be real. None
of these tests indicated a spurious detection, i.e., they are all consistent
with our result being a true detection of cycle-to-cycle differences in the
angular variability (see the online Appendix\,\ref{app:A} for details).
Specifically, we tested for:
\begin{enumerate}
  \item the impact of the slightly different quadruplet baselines used at the
  two epochs near maximal diameter, since one station differed between run A
  (I1) and run C (J3), cf. Tab.\,\ref{tab:vlti:quality};
  \item the impact of an asymmetric circumstellar envelope
  \citep[CSE;][]{2009A&A...498..425K}. To this end, we discarded short
  baselines ($V^2 > 0.5$) and re-determined diameters;
  \item the impact of the wavelength calibration. We investigate the
  intra-run stability for run C using HD\,74088 and the intra-run stability
  (run B to C) using HD\,81101. Since no standard stars were observed during
  run A, we unfortunately cannot demonstrate the stability of the inferred
  diameters over the full pulsation cycle. However, UD diameters of HD\,81101
  exhibit no systematic difference between run B and C;
  \item the impact of the fitting routine applied. We determine
  $\Theta_{\rm{UD}}$ and $\Theta_{\rm{LD,lin}}$ using {\tt LITpro} and
  $\Theta_{\rm{SATLAS}}$ using an independent python routine as well as {\tt
  CANDID}'s \citep{2015A&A...579A..68G} functionality to determine diameters.
  While the face values of diameters determined with a given software may differ
  numerically, the clear cycle-to-cycle difference remains;
  \item We test for companions of \lc\ (cf. Sect.\,\ref{sec:disc:companion}) as
  well as calibrator/standard stars by inspecting closure phases; 
\end{enumerate}
Other possible sources of systematic effects are listed in Appendix \ref{app:A}.
Future observing runs will include additional standard star observations to
monitor the stability of inferred diameters.

A real cycle-to-cycle difference between the two consecutive maximal diameters
of \lc\ should lead to differences in brightness and color at the corresponding
pulsation phase, since luminosity, radius, and temperature are linked to each
other via the Stefan$-$Boltzmann law. We estimate the effect on both brightness
and color while adopting a fixed value for the other.
For an unchanged temperature, luminosity will vary by $\sim 5\%$, leading to a
difference of approximately $10$\,mmag in the bolometric magnitude.
For constant luminosity, the change in radius would lead to a decrease of $10 -
15$\,K in effective temperature, which will be challenging to detect
spectroscopically. However, such a difference is likely detectable
photometrically with the {\tt BRITE}
nano-satellites\footnote{\url{http://brite-constellation.at/}}, since it
corresponds to a color difference of approximately $\Delta(B-V) \approx
5$\,mmag.
Magnitude fluctuations on this order of magnitude have previously been reported
using photometry from the {\tt MOST} satellite \citep{2015MNRAS.446.4008E} and,
at a lower level, for V1154\,Cygni with {\it Kepler}
\citep{2012MNRAS.425.1312D}, while \citet{2015arXiv150807639P} report only
tentative detections of modulation based on {\it CoRoT} photometry of seven
Cepheids. This illustrates the difficulty of detecting 
cycle-to-cycle differences, even with high-quality instruments due to the need
for high instrumental stability over the time-scales of at least two pulsation
cycles (more than two months for \lc).

\subsection{Modulated Spectroscopic Variability}\label{sec:spec}
\subsubsection{Radial Velocities}\label{sec:spec:RVs}

\lc\ was among the targets for which \citet{2014A&A...566L..10A} discovered the
modulated nature of Cepheid RV curves.
Our new (2015) \Coralie\ data presented here are approximately five times more
precise than the 2014 data, owing mainly to contemporaneous RV drift corrections and to a lesser
extent to an instrumental upgrade, cf. Sect.\,\ref{sec:obs:coralie}. As a
result, our new data are particularly sensitive to RV curve modulation.

\begin{figure*}
\centering
\includegraphics{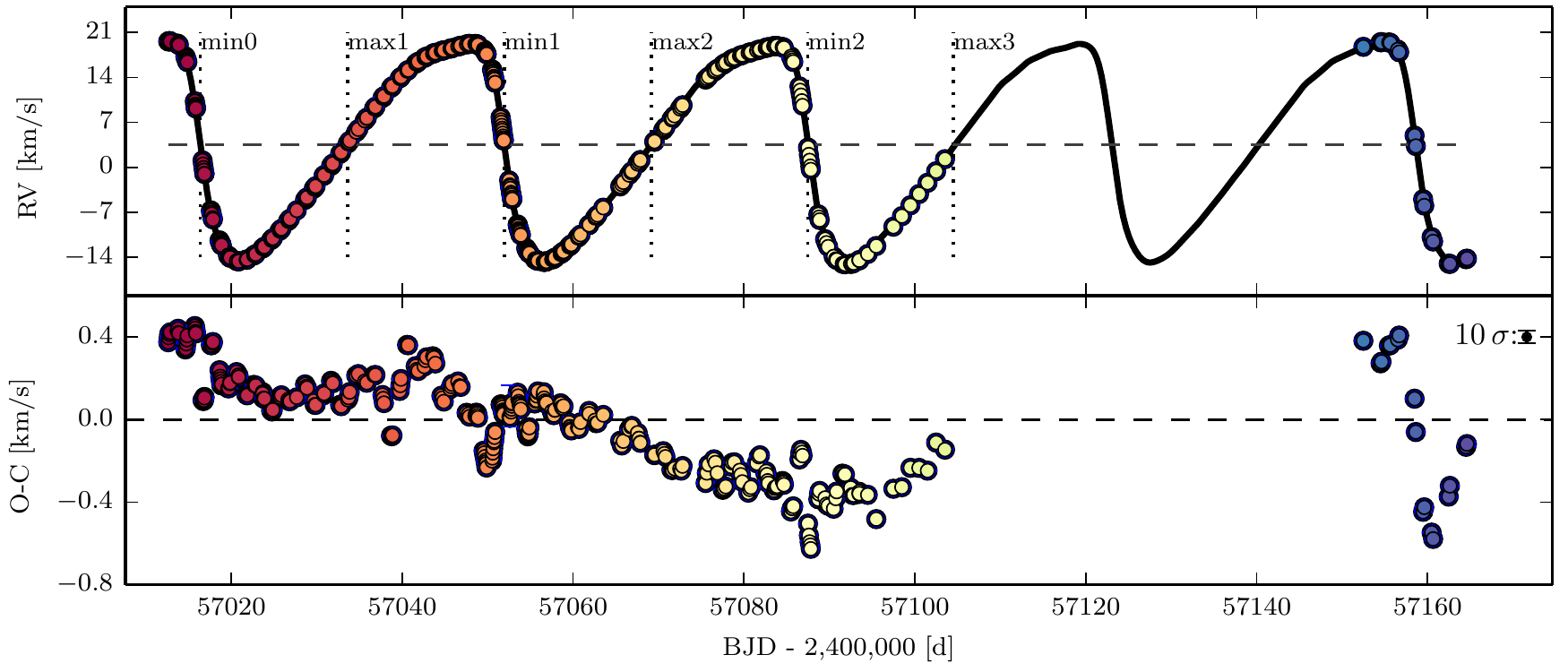}
\caption{The 2015 RV curve recorded for \lc\ against observation date. The top
shows the barycentric RVs determined via the cross-correlation technique, with 
a color coding that traces the observation date (same in other related figures).
Times near extremal diameters are indicated by min0, max1, min1, max2, min2, to inform
the discussion in the text. A 12th-order Fourier series fit is shown as a black
solid line, assuming pulsation period $P = 35.5578$\,d. The 10-fold median
uncertainty is shown in the top right of the bottom panel. The bottom panel
shows the residuals (data minus Fourier series fit), indicating a linear
decrease in average RV during the  continuously sampled part of the RV curve. 
This trend is not reproduced by the later, separate epoch observed near BJD
$2457160$.}
\label{fig:spec:2015rvcurve}
\end{figure*}

\begin{figure*}
\centering
\includegraphics{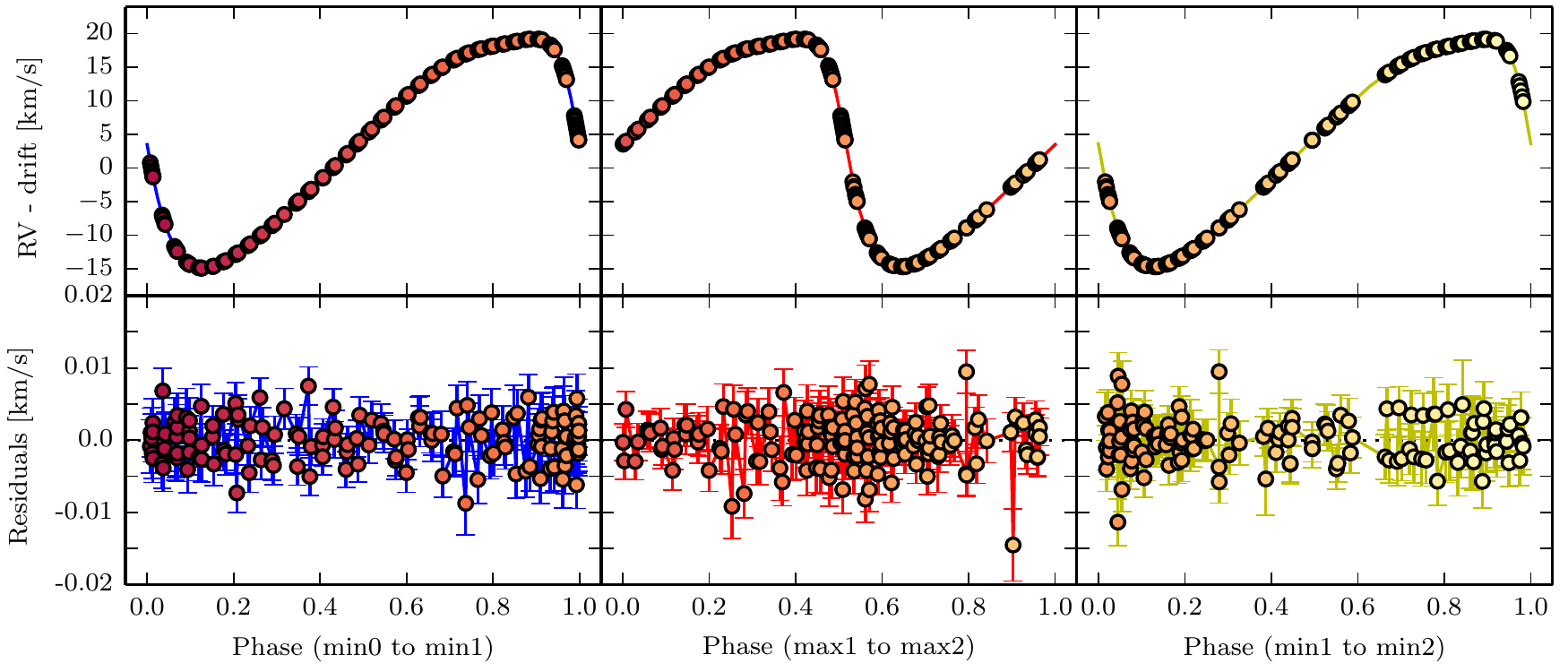}
\caption{The 2015 RV curve of \lc\ split into three separate parts, from min0
to min1 (left-hand panels), from max1 to max2 (center panels), and from min1 to min2
(right-hand panels).
The linear drift shown in Fig.\,\ref{fig:spec:2015rvcurve} was removed to
determine times when the RV curve intersects the average velocity. 
Top panels show the RV data minus the
drift as function of phase (within a given cycle). Bottom panels show the
residuals of these data minus a cubic spline fit (shown as solid line in 
upper panels). The rms of the three residual panels is $3.0\,$\ms, $3.3\,$\ms,
and $3.0\,$\ms (the median RV uncertainty is $3.14$\,\ms ) , respectively.}
\label{fig:spec:2015rvcurve_extremasep}
\end{figure*}

Figure\,\ref{fig:spec:2015rvcurve} shows the new \Coralie\ RV data obtained for
\lc. The top panel shows the RV with respect to the Solar system
barycenter against observation date and illustrates the unprecedented time
sampling achieved over a period of 3 full months. We label epochs of extremal
radii (min0, max1, min1, max2, min2, max3), which we use for timing purposes in
this work.
We compute residuals as RV minus a best-fitting 12-th order Fourier series
\citep[fitted following][]{2013MNRAS.434.2238A}. These residuals exhibit a
linear trend over the duration of the continuous monitoring campaign.
It is common to interpret such a linear variation in average RV as evidence for
spectroscopic binarity \citep[e.g.][]{2015ApJ...804..144A}. We therefore
obtained additional measurements in May\,2015 to test whether the trend would
continue. As the residuals demonstrate, the additional (May) data are not
consistent with a spectroscopic companion and show that the linear trend had
stopped. We note that changes in pulsation period cannot explain this
behavior. The recorded spectra exhibit cycle-to-cycle differences in the line
profile variations that appear to be the leading origin of RV curve modulation,
consistent with the interpretation that the apparent linear trend is not due
to an unseen companion. 
An in-depth discussion of this effect will be presented separately (Anderson, in
prep.).

To precisely determine epochs near extremal radii, knowledge of the intersection
points of the RV curve and its average is required. Since the average
velocity, usually referred to as $v_\gamma$, is {\it apparently} changing over the
duration of our contemporaneous interferometric and spectroscopic observing
campaign, we deem it appropriate to account for a time-variable $v_\gamma$ when
determining times of extremal radii. To this end, we represent the RV data
before epoch max3 by a 12-th order Fourier series and a linear trend
(slope $m_{\rm{drift}} = -8.0\,$\ms\,d$^{-1}$). We subtract this trend from the
RV data and from thereon treat the time series as having a constant $v_\gamma = 3.500$\,\kms .

We determine epochs of extremal radius from a cubic B-spline representation of
the RV curve sampled at intervals of $10^{-6}$\,d ($0.086$\,s), which corresponds to
the timing precision with which the spectra are recorded.
Table\,\ref{tab:RVs:timing} lists the timings of the extremal epochs
thus determined. We note that subtracting the linear trend from the RV curve
before determining epochs of extremal radius yields significantly
smaller scatter in the cycle-to-cycle behavior of the pulsation period.

\begin{table}
\begin{tabular}{lr}
Extremum & BJD - 2\,400\,000 \\
 & [d] \\
\hline
min0 &  57016.386006 \\
max1 &  57033.608096    \\
min1 &  57051.957192   \\
max2 &  57069.171281   \\
min2 &  57087.491377   \\
max3$^{\dagger}$ & 57104.522379 \\
\hline 
\end{tabular}
\caption{Epochs of extremal radius determined from the (trend-corrected) RV
curve. $^\dagger$ indicates that the epoch for max3 was determined by linear
extrapolation over less than 1 d.}
\label{tab:RVs:timing}
\end{table}

Figure\,\ref{fig:spec:2015rvcurve_extremasep} shows the RV curve of individual
pulsation cycles between extrema min0 and max2, defined as running from either
minimal radius to minimal radius or maximal to maximal radius over the duration
of the VLTI campaign (ran from max1 to max2). The residuals in the lower panel
have rms of approximately $3$\,\ms\ and show the excellent fit quality achieved
by the spline model\footnote{We chose spline models here, since Fourier series
models exhibit ringing that overall yields a slightly less credible
representation of the RV curve. While this effect is small, with a
typical rms of $6$\,\ms\ when using Fourier series, we prefer the spline
representation.}.

\begin{figure}
\centering
\includegraphics{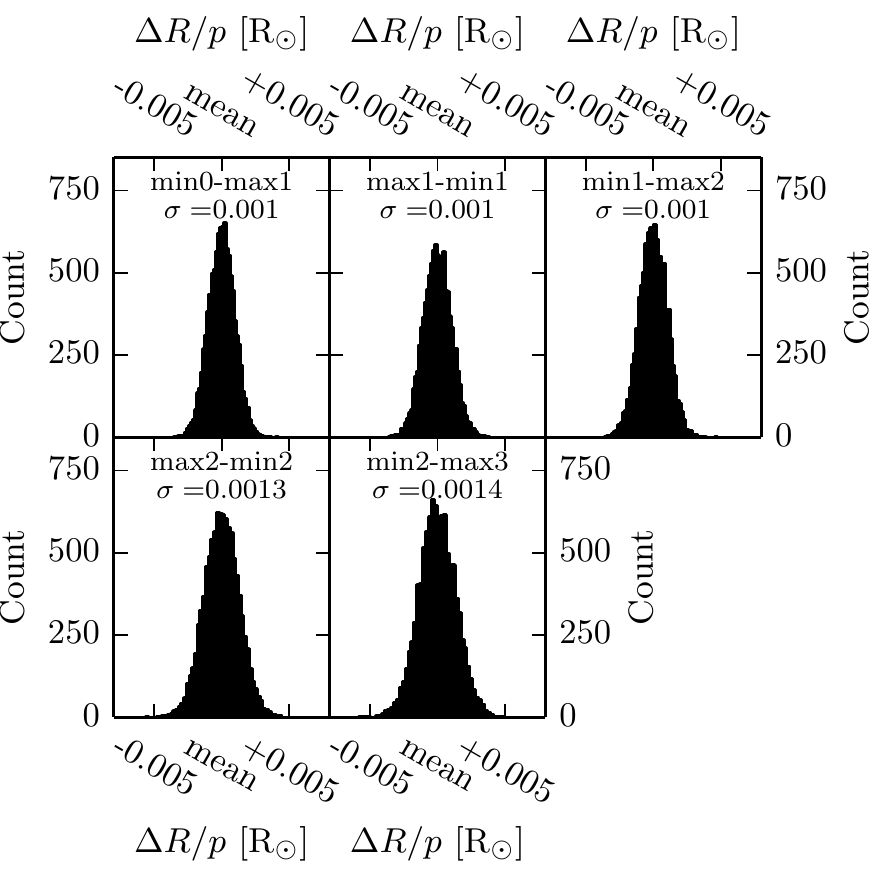}
\caption{Distributions of $\drp$ from $10000$ Monte Carlo trials, centered
on the respective means listed in Tab.\,\ref{tab:RVs:amps}. Standard deviations
are indicated inside each panel.}
\label{fig:RVs:MonteCarlo}
\end{figure}

We list the most relevant information regarding RV curve modulation in
Table\,\ref{tab:RVs:amps}. We provide durations of consecutive
expansion/contraction cycles, their RV amplitudes\footnote{We refer to expansion
and contraction cycles separately, since these enter BW
distances via $\Delta R$ and $\Delta \Theta$. The amplitudes of expansion and
contraction cycles have to be added to obtain the total peak-to-peak amplitude of a full pulsation cycle.}, as well
as the integral of the RV curve, i.e., $\drp$ (Eq.\,\ref{eq:deltaR}), determined
using the timings from Tab.\,\ref{tab:RVs:timing}. We determine $\Delta t$ and
$\drp$ as well as their uncertainties by means of a Monte Carlo experiment with
$10\,000$ trials. We show the distributions for $\drp$ thus determined in
Fig.\,\ref{fig:RVs:MonteCarlo}, centered on the mean of the distribution. We
determine RV amplitudes from the spline fit and list as a conservative
uncertainty estimate the rms of the residuals rounded up to the next $5$\,\ms .

\begin{table*}
\centering
\begin{tabular}{@{}lllllllll@{}}
\hline
From & Mean JD-2.4M & $\Delta t$ & $\sigma(\Delta t)$ & 
$N_{\rm{meas}}$ & $A_{\rm{RV}}$ & $\sigma(A_{\rm{RV}})$ & $\Delta R / p$ & $\sigma(\Delta R / p)$ \\
 & [d] & [d] & [d] & & [\kms] & [\kms] & [\Rsol] & [\Rsol] \\
\hline \smallskip
min0 to max1 & 57023.222701 & 17.2211 & 0.0014 & 66 & 18.391 & 0.005 & -24.0306 & 0.0095 \\\smallskip
max1 to min1 & 57045.413951 & 18.3521 & 0.0014 & 81 & 15.704 & 0.005 & 23.9752 & 0.0010 \\\smallskip 
min1 to max2 & 57058.037466 & 17.2118 & 0.0013 & 85 & 18.181 & 0.005 & -23.7519 & 0.0010 \\\smallskip 
max2 to min2 & 57079.597885 & 18.3197 & 0.0013 & 57 & 15.651 & 0.005 & 23.6587 &
0.0013 \\\smallskip
min2 to max3$^\dagger$ & 57092.907763 & 17.07 & 0.015 & 32 & 18.297 & 0.010 &
-23.7910 & 0.0020 \\
\hline
\end{tabular}
\caption{RV (semi-)amplitudes $A_{\rm{RV}}$ and RV curve integrals $\drp$
determined from new \Coralie\ data. The mean Julian date is that of the
observations between consecutive extrema. The duration between extrema $\Delta
t$ and $\drp$ as well as their uncertainties are determined from a 
Monte Carlo analysis with 10000 repetitions. RV (semi-)amplitudes
$A_{\rm{RV}}$ are estimated from a spline fit (cf.
Fig.\,\ref{fig:spec:2015rvcurve} with a fixed adopted uncertainty of $5$\,\ms,
for an rms of approximately $3$\,\ms .
We used $R_\odot = 696\,342 \pm 65$\,km \citep{2012ApJ...750..135E}. The
$^{\dagger}$ indicates that a linear extrapolation was used to determine
the values for the last epoch (min2-max3), see Sect.\,\ref{sec:spec}.}
\label{tab:RVs:amps}
\end{table*}

The values of $\Delta t$ in Tab.\,\ref{tab:RVs:amps} 
indicate that \lc\  contracts during $\sim 51.5\%$ of its pulsation cycle.
The durations also exhibit clear cycle-to-cycle variations. Such `jitter' in
pulsation periods has also been identified using space-based photometry of
certain short period Cepheids, see \citet{2012MNRAS.425.1312D} and
\citet{2015MNRAS.446.4008E}. We caution that the duration of the last expansion
cycle (min2 to max3) appears to be significantly affected by the
upturning RV residuals in Fig.\,\ref{fig:spec:2015rvcurve}, where the linear
drift adds bias. 

\begin{figure}
\centering
\includegraphics{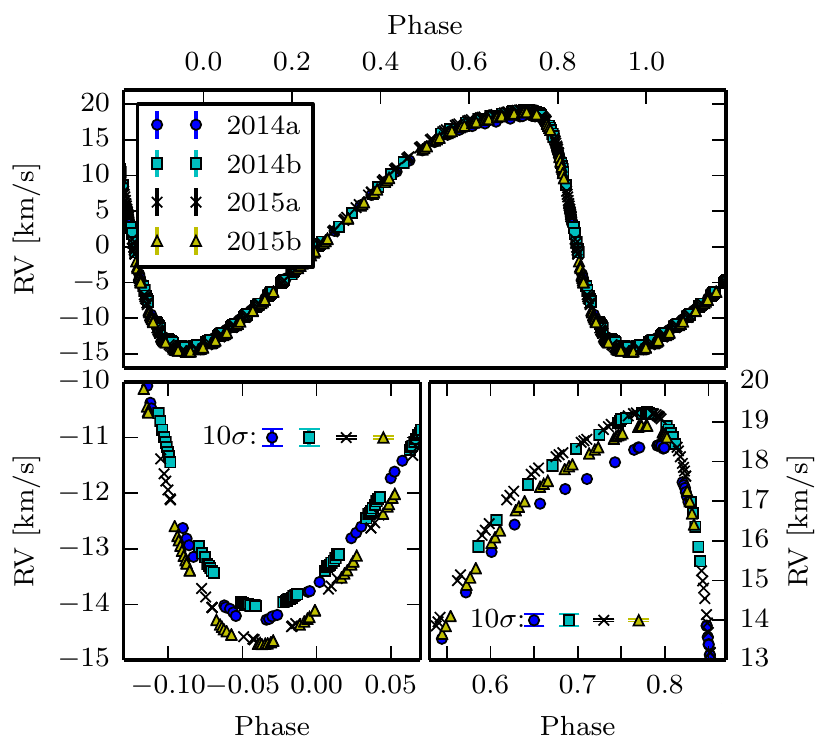}
\caption{Comparison between 2014 and 2015 behavior of RV curve modulation. Data
from two consecutive pulsation cycles from 2014 and 2015. 2014 data
are shown as blue circles and cyan squares, data from 2015 as blue crosses
(min0-min1) and yellow dots (min1-min2). In 2014, overall RV curve
amplitudes were lower, but cycle-to-cycle differences were greater than in
2015. 10-fold median uncertainties shown in bottom panels.}
\label{fig:RVs:2014vs2015}
\end{figure}

In addition to period jitter, Tab.\,\ref{tab:RVs:amps} also clearly shows
varying pulsation amplitudes, albeit at a much lower level than the one shown in
\citet[$\Delta A_{v_r} = 0.51$\,\kms\ between two consecutive
cycles]{2014A&A...566L..10A}. We illustrate this by plotting two well-sampled
sets of two consecutive pulsation cycles from 2014 and 2015 in
Fig.\,\ref{fig:RVs:2014vs2015}, demonstrating the irregular and unpredictable
nature of RV curve modulation. We point out that RV zero-point offsets (of order
$0.015\,$\kms) due to the instrumental upgrade are not responsible for this
difference in amplitude. While all values of $\drp$ determined from 2015 data
are consistently larger than the ones from the year before, we note that the
cycle-to-cycle variations were weaker in 2015. 

Summing up the values for $\drp$ determined in 2015, we find that RVs indicate
an overall {\it decrease} in stellar radius of $0.223 \pm 0.001$\,\Rsol\ between
the epochs with contemporaneous interferometric measurements (max1 and max2),
cf. Sect.\,\ref{sec:AngDiam}, in contrast with the interferometric results. 
Taken at face value, the modulated RV curve indicates\footnote{assuming a
``true" $d=497.5\,$pc \citep[value from][]{2007AJ....133.1810B} and $p = 1.22$
from Breitfelder et al. (in preparation)}
a {\it decrease} of $4.2\,\mu$as between
epochs max1 and max2, i.e., an effect of different sign and amplitude compared
to the angular diameter variability investigated above. One way to reconcile
such a difference is by introducing a dependence of projection factors on
pulsation cycles, cf. Sect.\,\ref{sec:discuss:BW}.

\section{Discussion}\label{sec:discuss}
We now discuss these findings. First, we investigate whether the diameters we
determine for \lc\ may be biassed due to an unseen companion, finding no
indication of this. Then, we discuss the implications of modulated angular
and linear variability on BW distances.

\subsection{Possible companion stars to \lc}\label{sec:disc:companion}
Based on the interferometric observations, we determine a detection limit of
$0.15\%$, i.e., $\Delta m_H \approx 7$\,mag, for any companions within $\sim
50\,$mas of \lc\ using the {\tt CANDID} tool \citep{2015A&A...579A..68G}. At a
distance of $\sim 500$\,pc this excludes bright companions with relative
semimajor axis $a_{\rm{rel}} \lesssim 25$\,AU.

Based on this detection limit and Geneva stellar evolution models
\citep{2012A&A...537A.146E,2013A&A...553A..24G,2014A&A...564A.100A} of solar
metallicity and average rotation, we estimate that \lc\ does not have a
companion with mass greater than $5$\,\Msol (assuming $9\,$\Msol\ for \lc).
The models predict $M_H \approx -7.6$\,mag for such a Cepheid on the third
crossing of the instability strip, which is consistent with observations of
rates of period change (Breitfelder et al. in prep.) and the absolute magnitude
measured by \citet{2007AJ....133.1810B}.
This provides an age estimate of approximately $35$\,Myr, which we use to inform
the upper mass limit for a main sequence (MS) companion.

Possible companions with contrast ratios higher than $0.05\%$ could in principle
bias our result, while being undetectable in the interferometric data.
This contrast ratio corresponds to a $\sim 3.5\,$\Msol\ MS star.
Adopting an orbit with $a_{\rm{rel}} = 25\,$AU (the detection window for {\tt
CANDID}), and a circular orbit, we find that such a hypothetical companion would
have a semi-amplitude of $> 1\,$\kms\ for inclination $i > 10\,$ deg
(e.g. $2\,$\kms\ at $i \sim 20$\,deg) and orbital period of $\sim 35\,$ yr.
However, such a companion would be unlikely to bias the diameter estimate.
Closer-in orbits would be more easily detectable due to higher RV semi-amplitude
and shorter orbital periods. For instance, a semimajor axis of $3.5$\,AU ($\sim
5$ times the radius of \lc) would have $P_{\rm{orb}} \sim 1.9\,$yr, and orbital
RV semi-amplitude in excess of $1\,$\kms\ for almost all possible inclinations
($i > 4$\,deg.)

If one were to adopt the residual drift in RVs seen in
Fig.\,\ref{fig:spec:2015rvcurve} as a sign of spectroscopic binarity, then this
would yield an estimate of the orbital period in the range of approximately
$160$\,d. We estimate a minimum mass of approximately
$0.05\,$\Msol\ for an edge-on orbit with this period and $K_1 \sim 400\,$\ms.
For a companion star of more significant mass, the orbit 
would have to be seen nearly face on for this not to cause
enormous RV variations.
Within the range of the residuals, only inclinations $< 1$\, deg would be
allowed. However, such an orbit would result in significant tidal forces that
would significantly distort \lc, leading to hotter polar regions. Due to the
(hypothetical) low inclination, one would thus expect \lc\ as an abnormally hot
Cepheid. Contrastingly, \lc\ is among the coolest known Cepheids, i.e., this
prediction is inconsistent with the available evidence.

It is thus highly unlikely that our diameter measurements of \lc\ are biased by
an unseen companion. Low-mass companions on long-period orbits remain possible,
however.  {\it Gaia}'s ongoing observations of \lc\ will be useful for further
exploring the parameter space of possible companions to \lc .

\subsection{Importance for Baade-Wesselink distances}\label{sec:discuss:BW}
As described above, we find tentative evidence for \lc 's different maximal
angular diameters measured from two consecutive pulsation cycles. 
Contemporaneously, we have confirmed its modulated variability in RV.
While interferometric measurements suggest (at face value) an {\it
increase} in maximal diameter from the first to the second maximum (by approximately
$22.5\,\mu$as for limb darkened diameters), the modulated RV curve suggests a
variation with opposite sign and smaller amplitude.

If confirmed by further interferometric measurements, this discrepancy could be 
explained by differences in the motion of {\it gas} layers (traced by RVs)
and {\it optical} layers (interferometry traces motion of the continuum).
Following \citet{2007A&A...471..661N}, this difference enters into the
definition of the projection factor, which is used to translate
measured line-of-sight velocities to pulsation velocities, cf. 
the factor $f_{o-g}$ in Eq.\,\ref{eq:pfactor}. Our result would thus imply 
a time-dependence in the differential motion between the optical and gas motions,
rendering $f_{o-g}$ time-dependent, i.e., differing between one
contraction or expansion cycle and the following expansion or contraction. 

To quantify the possible difference in $p-$factor caused by
differentially moving layers, we consider the following scenarios in which we
determine $p$ using (a ``true'' distance of) $d= 497.5\,$pc \citep[the
uncertainty of approximately $10\%$ is neglected for this thought
experiment]{2007AJ....133.1810B}, and the three possible combinations of $\Delta
\Theta$ and $\Delta R/p$ measured contemporaneously. We list the results in
Tab.\,\ref{tab:disc:pfactors}.
\begin{table}
\centering
\begin{tabular}{lrrrrr}
combination & $\Theta_{\rm{LD,max}}$ & $\Theta_{\rm{LD,min}}$ & $\Delta \Theta$
& $\drp$ & $p$ \\
 & [mas] & [mas] & [mas] & [\Rsol] & \\
\hline
max1 to min1 & 3.2482 & 2.6905 & 0.5577 & 23.9752 & 1.24 \\
min1 to max2 & 3.2707 & 2.6905 & 0.5802 & 23.7519 & 1.31 \\
max1 to max2 &  & &  1.1379 & 47.7271 & 1.2741 \\
\hline
\end{tabular}
\caption{$p-$factors calculated separately for contraction and expansion, and
for the full pulsation cycle (sum of the two) based on our fully contemporaneous
data set. We here adopt $d = 497.5$\,pc as \emph{true} distance to showcase the
impact of modulated variability on $p$-factors. We use the limb-darkened diameters based on
SATLAS models, cf. Tab.\,\ref{tab:vlti:diams}.
The accuracy of $p-$factors is currently limited by errors on distance
($\sim 10\%$).}
\label{tab:disc:pfactors}
\end{table}

These examples illustrate how cycle-to-cycle variations in $\Delta \Theta$ and
the integrated RV curve lead to differences in $p$ of up to $5\%$, even between
successive contractions and expansions. This significant difference must be
overcome to enable BW distances accurate to $1\%$, since BW distances depend
linearly on $p$.  However, cycle-to-cycle variations in \lc\ thus far appear to
be stochastic (based on the 2014 and 2015 RV data). It therefore seems likely
that such effects largely cancel out when using data covering a larger temporal
baseline. This is the approach used in analyses based on the {\tt SPIPS} code
(M\'erand et al. in press; Breitfelder et al. in prep.). 

If not somehow avoided or corrected for,
the modulated variability of \lc\ and other Cepheids could lead to an
increased scatter of PLRs calibrated using BW distances. Decreasing the
scatter of PLRs is, however, crucial for separating luminosity
differences due to line-of-sight and metallicity effects in Cepheids located
in the Magellanic clouds.

\section{Conclusions}\label{sec:conclude}
We have carried out an unprecedented 3-month-long interferometric
(VLTI/\Pionier) and spectroscopic (\Coralie ) monitoring campaign aimed
at characterizing the modulated variability of the long-period Cepheid \lc . 

The following summarizes our key results.
\begin{enumerate}
  \item We find the first tentative evidence of 
  cycle-to-cycle differences in the {\it angular} diameter variability of a pulsating star. The two 
  maximal (limb-darkened) angular diameters determined from two consecutive
  epochs spanning maximal diameters differ by $22.4 \pm 1.4 \mathrm{(stat.)}
  \,\mu$as. Our results suggest the second maximum to subtend a larger angle.
  While certain systematic effects could render this
  result spurious, none of the tests that we were able to carry out indicated
  such bias.
  \item If real, this difference in maximal angular diameter can lead
  to photometric variations at this pulsation phase on the order of 
  $\sim 10$\,mmag in bolometric magnitude, or $5$\,mmag in $(B-V)$ color.
  Fluctuations on this order have been detected in short-period Cepheids by
  \citet{2012MNRAS.425.1312D} and \citet{2015MNRAS.446.4008E}.
  \item We confirm the presence of RV curve modulation reported by
  \citet{2014A&A...566L..10A}. Our new observations have approximately
  five times higher precision. We find a lower level of RV curve modulation than
  in 2014, although overall RV amplitudes are larger, indicating
  an irregular modulation. We find that RV curve modulation can under certain
  circumstances be misinterpreted as evidence for spectroscopic binarity.
  \item Our results suggest that angular diameter and
  linear radius variations are modulated at different amplitudes and (in
  this case) with opposite sign.
  Such behavior is likely indicative of the different motion
  undergone by the {\it optical} continuum traced interferometrically and the
  {\it gas} traced via RVs. This different motion enters the definition of the
  projection factor as a factor $f_{o-g}$,  see \citet{2007A&A...471..661N} and
  Eq.\,\ref{eq:pfactor}. Our result thus suggests that $f_{o-g}$ is
  time-dependent, varying by approximately $5\%$ between successive expansion and contraction cycles. The
  irregular behavior of the RV curve modulation suggests a complex
  time-dependence of $p$-factors. 
  \item We use interferometric measurements to set an upper limit of
  $5\,$\Msol\ for any potential companion stars. 
\end{enumerate}

Our detailed spectro-interferometric investigation reveals previously hidden
complexities of Cepheid pulsations that open a new window for understanding the
variability of classical Cepheids. Additional observations are
required to investigate the long-term behavior of the modulated variability as
well as any possible periodicity. To this end, 
high-quality interferometric, multi-band photometric, and
high-precision velocimetric data are needed.

Greater angular resolution, enabled by larger interferometric baselines
such as those offered by the {\it CHARA} array or high-precision instruments
such as the future {\it GRAVITY} instrument at ESO VLTI provide access to
expanding this kind of study to additional Cepheids. High-quality photometry
will soon be provided by the {\it BRITE} nano-satellites. Optical spectrographs
capable of delivering \ms\ precision are becoming more common thanks to the
developments driven by the search for extra-solar planets. Finally, {\it Gaia}
and {\it Hubble Space Telescope} \citep{2014ApJ...785..161R} will soon deliver Cepheid parallaxes
of unprecedented accuracy. This fortuitous combination of instruments delivering
unprecedented data quality will enable significant improvements for
Baade-Wesselink distances in the near future, which will enable detailed
investigations of the effect of chemical composition on the period-luminosity
relation.

\section*{Acknowledgments}
  We thank the anonymous referee for her/his report. 
  We acknowledge the Euler team at Geneva Observatory and La Silla Observatory
  for ensuring smooth operations and the Geneva exoplanet group for providing
  support.
  We thank (in chronological order) A.H.M.J.\ Triaud, S.\ P\'eretti, J.\ Cerda,
  D.\ Martin, V.\ M\'egevand, L.\ Weber, J.\ Jenkins, and V.\ Bonvin
  for carrying out observations as part of the RV monitoring campaign. We also
  thank the director of Geneva Observatory, S.\ Udry for agreeing to such a
  dedicated, long-term effort.    
  We greatly appreciated the friendly and competent assistance by
  ESO staff at ESO La Silla and Paranal Observatories. 

RIA acknowledges funding from the Swiss National Science Foundation.
PK, AM, JB, and AG acknowledge financial support from the ``Programme
  National de Physique Stellaire" (PNPS) of CNRS/INSU, France.
PK and AG acknowledge support of the French-Chilean exchange program
ECOS-Sud/CONICYT. 
AG acknowledges support from FONDECYT grant 3130361.
This research received the support of PHASE, the partnership between ONERA,
Observatoire de Paris, CNRS and University Denis Diderot Paris 7. 
The Euler telescope is supported by the Swiss National Science Foundation. 

This research has made use of the following services provided Jean-Marie
Mariotti Center services \texttt{Aspro}\footnote{Available at
\url{http://www.jmmc.fr/aspro}} and \texttt{SearchCal}\footnote{Available at
\url{http://www.jmmc.fr/searchcal}}, co-developed by FIZEAU and LAOG/IPAG, as
well as of the CDS Astronomical Databases
SIMBAD and VizieR catalogue access tool\footnote{Available at
\url{http://cdsweb.u-strasbg.fr/}} and NASA's Astrophysics Data System.

%-------------------------------------------------------------------

\bibliography{Bib_mine,Bib_modulation,Bib_Cepheids,Bib_Spectropolarimetry,Bib_StellarRotation,Bib_general,Bib_Spectroscopy,Bib_interferometry,RotatingCepBibTexRefs,Bib_StellarEvolution}

\hyphenation{Post-Script Sprin-ger}
\begin{thebibliography}{}
\makeatletter
\relax
\def\mn@urlcharsother{\let\do\@makeother \do\$\do\&\do\#\do\^\do\_\do\%\do\~}
\def\mn@doi{\begingroup\mn@urlcharsother \@ifnextchar [ {\mn@doi@}
  {\mn@doi@[]}}
\def\mn@doi@[#1]#2{\def\@tempa{#1}\ifx\@tempa\@empty \href
  {http://dx.doi.org/#2} {doi:#2}\else \href {http://dx.doi.org/#2} {#1}\fi
  \endgroup}
\def\mn@eprint#1#2{\mn@eprint@#1:#2::\@nil}
\def\mn@eprint@arXiv#1{\href {http://arxiv.org/abs/#1} {{\tt arXiv:#1}}}
\def\mn@eprint@dblp#1{\href {http://dblp.uni-trier.de/rec/bibtex/#1.xml}
  {dblp:#1}}
\def\mn@eprint@#1:#2:#3:#4\@nil{\def\@tempa {#1}\def\@tempb {#2}\def\@tempc
  {#3}\ifx \@tempc \@empty \let \@tempc \@tempb \let \@tempb \@tempa \fi \ifx
  \@tempb \@empty \def\@tempb {arXiv}\fi \@ifundefined
  {mn@eprint@\@tempb}{\@tempb:\@tempc}{\expandafter \expandafter \csname
  mn@eprint@\@tempb\endcsname \expandafter{\@tempc}}}

\bibitem[\protect\citeauthoryear{{Anderson}}{{Anderson}}{2013}]{2013PhDT.......363A}
{Anderson} R.~I.,  2013, PhD thesis, Universit{\'e} de Gen{\`e}ve

\bibitem[\protect\citeauthoryear{{Anderson}}{{Anderson}}{2014}]{2014A&A...566L..10A}
{Anderson} R.~I.,  2014, \mn@doi [\aap] {10.1051/0004-6361/201423850}, \href
  {http://adsabs.harvard.edu/abs/2014A%26A...566L..10A} {566, L10}

\bibitem[\protect\citeauthoryear{{Anderson}, {Eyer}  \& {Mowlavi}}{{Anderson}
  et~al.}{2013}]{2013MNRAS.434.2238A}
{Anderson} R.~I.,  {Eyer} L.,   {Mowlavi} N.,  2013, \mn@doi [\mnras]
  {10.1093/mnras/stt1160}, \href
  {http://adsabs.harvard.edu/abs/2013MNRAS.434.2238A} {434, 2238}

\bibitem[\protect\citeauthoryear{{Anderson}, {Ekstr{\"o}m}, {Georgy}, {Meynet},
  {Mowlavi}  \& {Eyer}}{{Anderson} et~al.}{2014}]{2014A&A...564A.100A}
{Anderson} R.~I.,  {Ekstr{\"o}m} S.,  {Georgy} C.,  {Meynet} G.,  {Mowlavi} N.,
    {Eyer} L.,  2014, \mn@doi [\aap] {10.1051/0004-6361/201322988}, \href
  {http://adsabs.harvard.edu/abs/2014A%26A...564A.100A} {564, A100}

\bibitem[\protect\citeauthoryear{{Anderson}, {Sahlmann}, {Holl}, {Eyer},
  {Palaversa}, {Mowlavi}, {S{\"u}veges}  \& {Roelens}}{{Anderson}
  et~al.}{2015}]{2015ApJ...804..144A}
{Anderson} R.~I.,  {Sahlmann} J.,  {Holl} B.,  {Eyer} L.,  {Palaversa} L.,
  {Mowlavi} N.,  {S{\"u}veges} M.,   {Roelens} M.,  2015, \mn@doi [\apj]
  {10.1088/0004-637X/804/2/144}, \href
  {http://adsabs.harvard.edu/abs/2015ApJ...804..144A} {804, 144}

\bibitem[\protect\citeauthoryear{{Baade}}{{Baade}}{1926}]{1926AN....228..359B}
{Baade} W.,  1926, Astronomische Nachrichten, \href
  {http://adsabs.harvard.edu/abs/1926AN....228..359B} {228, 359}

\bibitem[\protect\citeauthoryear{{Baranne} et~al.,}{{Baranne}
  et~al.}{1996}]{1996A&AS..119..373B}
{Baranne} A.,  et~al., 1996, \aaps, \href
  {http://adsabs.harvard.edu/abs/1996A%26AS..119..373B} {119, 373}

\bibitem[\protect\citeauthoryear{{Becker}}{{Becker}}{1940}]{1940ZA.....19..289B}
{Becker} W.,  1940, \zap, \href
  {http://adsabs.harvard.edu/abs/1940ZA.....19..289B} {19, 289}

\bibitem[\protect\citeauthoryear{{Benedict} et~al.,}{{Benedict}
  et~al.}{2007}]{2007AJ....133.1810B}
{Benedict} G.~F.,  et~al., 2007, \mn@doi [\aj] {10.1086/511980}, \href
  {http://adsabs.harvard.edu/abs/2007AJ....133.1810B} {133, 1810}

\bibitem[\protect\citeauthoryear{{Bonneau} et~al.,}{{Bonneau}
  et~al.}{2006}]{2006A&A...456..789B}
{Bonneau} D.,  et~al., 2006, \mn@doi [\aap] {10.1051/0004-6361:20054469}, \href
  {http://cdsads.u-strasbg.fr/abs/2006A%26A...456..789B} {456, 789}

\bibitem[\protect\citeauthoryear{{Bonneau}, {Delfosse}, {Mourard}, {Lafrasse},
  {Mella}, {Cetre}, {Clausse}  \& {Zins}}{{Bonneau}
  et~al.}{2011}]{2011A&A...535A..53B}
{Bonneau} D.,  {Delfosse} X.,  {Mourard} D.,  {Lafrasse} S.,  {Mella} G.,
  {Cetre} S.,  {Clausse} J.-M.,   {Zins} G.,  2011, \mn@doi [\aap]
  {10.1051/0004-6361/201015124}, \href
  {http://cdsads.u-strasbg.fr/abs/2011A%26A...535A..53B} {535, A53}

\bibitem[\protect\citeauthoryear{{Bord{\'e}}, {Coud{\'e} du Foresto}, {Chagnon}
   \& {Perrin}}{{Bord{\'e}} et~al.}{2002}]{2002A&A...393..183B}
{Bord{\'e}} P.,  {Coud{\'e} du Foresto} V.,  {Chagnon} G.,   {Perrin} G.,
  2002, \mn@doi [\aap] {10.1051/0004-6361:20021020}, \href
  {http://cdsads.u-strasbg.fr/abs/2002A%26A...393..183B} {393, 183}

\bibitem[\protect\citeauthoryear{{Bouchy}, {Pepe}  \& {Queloz}}{{Bouchy}
  et~al.}{2001}]{2001A&A...374..733B}
{Bouchy} F.,  {Pepe} F.,   {Queloz} D.,  2001, \mn@doi [\aap]
  {10.1051/0004-6361:20010730}, \href
  {http://adsabs.harvard.edu/abs/2001A%26A...374..733B} {374, 733}

\bibitem[\protect\citeauthoryear{{Bouchy}, {D{\'{\i}}az}, {H{\'e}brard},
  {Arnold}, {Boisse}, {Delfosse}, {Perruchot}  \& {Santerne}}{{Bouchy}
  et~al.}{2013}]{2013A&A...549A..49B}
{Bouchy} F.,  {D{\'{\i}}az} R.~F.,  {H{\'e}brard} G.,  {Arnold} L.,  {Boisse}
  I.,  {Delfosse} X.,  {Perruchot} S.,   {Santerne} A.,  2013, \mn@doi [\aap]
  {10.1051/0004-6361/201219979}, \href
  {http://adsabs.harvard.edu/abs/2013A%26A...549A..49B} {549, A49}

\bibitem[\protect\citeauthoryear{{Davis}, {Jacob}, {Robertson}, {Ireland},
  {North}, {Tango}  \& {Tuthill}}{{Davis} et~al.}{2009}]{2009MNRAS.394.1620D}
{Davis} J.,  {Jacob} A.~P.,  {Robertson} J.~G.,  {Ireland} M.~J.,  {North}
  J.~R.,  {Tango} W.~J.,   {Tuthill} P.~G.,  2009, \mn@doi [\mnras]
  {10.1111/j.1365-2966.2009.14433.x}, \href
  {http://adsabs.harvard.edu/abs/2009MNRAS.394.1620D} {394, 1620}

\bibitem[\protect\citeauthoryear{{Derekas} et~al.,}{{Derekas}
  et~al.}{2012}]{2012MNRAS.425.1312D}
{Derekas} A.,  et~al., 2012, \mn@doi [\mnras]
  {10.1111/j.1365-2966.2012.21538.x}, \href
  {http://esoads.eso.org/abs/2012MNRAS.425.1312D} {425, 1312}

\bibitem[\protect\citeauthoryear{{Ekstr{\"o}m} et~al.,}{{Ekstr{\"o}m}
  et~al.}{2012}]{2012A&A...537A.146E}
{Ekstr{\"o}m} S.,  et~al., 2012, \mn@doi [\aap] {10.1051/0004-6361/201117751},
  \href {http://adsabs.harvard.edu/abs/2012A%26A...537A.146E} {537, A146}

\bibitem[\protect\citeauthoryear{{Emilio}, {Kuhn}, {Bush}  \&
  {Scholl}}{{Emilio} et~al.}{2012}]{2012ApJ...750..135E}
{Emilio} M.,  {Kuhn} J.~R.,  {Bush} R.~I.,   {Scholl} I.~F.,  2012, \mn@doi
  [\apj] {10.1088/0004-637X/750/2/135}, \href
  {http://adsabs.harvard.edu/abs/2012ApJ...750..135E} {750, 135}

\bibitem[\protect\citeauthoryear{{Evans} et~al.,}{{Evans}
  et~al.}{2015}]{2015MNRAS.446.4008E}
{Evans} N.~R.,  et~al., 2015, \mn@doi [\mnras] {10.1093/mnras/stu2371}, \href
  {http://adsabs.harvard.edu/abs/2015MNRAS.446.4008E} {446, 4008}

\bibitem[\protect\citeauthoryear{{Feast} \& {Catchpole}}{{Feast} \&
  {Catchpole}}{1997}]{1997MNRAS.286L...1F}
{Feast} M.~W.,  {Catchpole} R.~M.,  1997, \mnras, \href
  {http://adsabs.harvard.edu/abs/1997MNRAS.286L...1F} {286, L1}

\bibitem[\protect\citeauthoryear{{Fernie}, {Evans}, {Beattie}  \&
  {Seager}}{{Fernie} et~al.}{1995}]{1995IBVS.4148....1F}
{Fernie} J.~D.,  {Evans} N.~R.,  {Beattie} B.,   {Seager} S.,  1995,
  Information Bulletin on Variable Stars, \href
  {http://adsabs.harvard.edu/abs/1995IBVS.4148....1F} {4148, 1}

\bibitem[\protect\citeauthoryear{{Fouqu{\'e}} \& {Gieren}}{{Fouqu{\'e}} \&
  {Gieren}}{1997}]{1997A&A...320..799F}
{Fouqu{\'e}} P.,  {Gieren} W.~P.,  1997, \aap, \href
  {http://esoads.eso.org/abs/1997A%26A...320..799F} {320, 799}

\bibitem[\protect\citeauthoryear{{Freedman}, {Madore}, {Scowcroft}, {Burns},
  {Monson}, {Persson}, {Seibert}  \& {Rigby}}{{Freedman}
  et~al.}{2012}]{2012ApJ...758...24F}
{Freedman} W.~L.,  {Madore} B.~F.,  {Scowcroft} V.,  {Burns} C.,  {Monson} A.,
  {Persson} S.~E.,  {Seibert} M.,   {Rigby} J.,  2012, \mn@doi [\apj]
  {10.1088/0004-637X/758/1/24}, \href
  {http://esoads.eso.org/abs/2012ApJ...758...24F} {758, 24}

\bibitem[\protect\citeauthoryear{{Gallenne} et~al.,}{{Gallenne}
  et~al.}{2015}]{2015A&A...579A..68G}
{Gallenne} A.,  et~al., 2015, \mn@doi [\aap] {10.1051/0004-6361/201525917},
  \href {http://adsabs.harvard.edu/abs/2015A%26A...579A..68G} {579, A68}

\bibitem[\protect\citeauthoryear{{Georgy}, {Ekstr{\"o}m}, {Granada}, {Meynet},
  {Mowlavi}, {Eggenberger}  \& {Maeder}}{{Georgy}
  et~al.}{2013}]{2013A&A...553A..24G}
{Georgy} C.,  {Ekstr{\"o}m} S.,  {Granada} A.,  {Meynet} G.,  {Mowlavi} N.,
  {Eggenberger} P.,   {Maeder} A.,  2013, \mn@doi [\aap]
  {10.1051/0004-6361/201220558}, \href
  {http://adsabs.harvard.edu/abs/2013A%26A...553A..24G} {553, A24}

\bibitem[\protect\citeauthoryear{{Groenewegen}}{{Groenewegen}}{2008}]{2008A&A...488...25G}
{Groenewegen} M.~A.~T.,  2008, \mn@doi [\aap] {10.1051/0004-6361:200809859},
  \href {http://adsabs.harvard.edu/abs/2008A%26A...488...25G} {488, 25}

\bibitem[\protect\citeauthoryear{{Groenewegen}}{{Groenewegen}}{2013}]{2013A&A...550A..70G}
{Groenewegen} M.~A.~T.,  2013, \mn@doi [\aap] {10.1051/0004-6361/201220446},
  \href {http://adsabs.harvard.edu/abs/2013A%26A...550A..70G} {550, A70}

\bibitem[\protect\citeauthoryear{{Hinshaw} et~al.,}{{Hinshaw}
  et~al.}{2013}]{2013ApJS..208...19H}
{Hinshaw} G.,  et~al., 2013, \mn@doi [\apjs] {10.1088/0067-0049/208/2/19},
  \href {http://adsabs.harvard.edu/abs/2013ApJS..208...19H} {208, 19}

\bibitem[\protect\citeauthoryear{{Hoffmann} \& {Macri}}{{Hoffmann} \&
  {Macri}}{2015}]{2015AJ....149..183H}
{Hoffmann} S.~L.,  {Macri} L.~M.,  2015, \mn@doi [\aj]
  {10.1088/0004-6256/149/6/183}, \href
  {http://adsabs.harvard.edu/abs/2015AJ....149..183H} {149, 183}

\bibitem[\protect\citeauthoryear{{Humphreys}, {Reid}, {Moran}, {Greenhill}  \&
  {Argon}}{{Humphreys} et~al.}{2013}]{2013ApJ...775...13H}
{Humphreys} E.~M.~L.,  {Reid} M.~J.,  {Moran} J.~M.,  {Greenhill} L.~J.,
  {Argon} A.~L.,  2013, \mn@doi [\apj] {10.1088/0004-637X/775/1/13}, \href
  {http://adsabs.harvard.edu/abs/2013ApJ...775...13H} {775, 13}

\bibitem[\protect\citeauthoryear{{Kervella}, {Coud{\'e} du Foresto}, {Perrin},
  {Sch{\"o}ller}, {Traub}  \& {Lacasse}}{{Kervella}
  et~al.}{2001}]{2001A&A...367..876K}
{Kervella} P.,  {Coud{\'e} du Foresto} V.,  {Perrin} G.,  {Sch{\"o}ller} M.,
  {Traub} W.~A.,   {Lacasse} M.~G.,  2001, \mn@doi [\aap]
  {10.1051/0004-6361:20000490}, \href
  {http://esoads.eso.org/abs/2001A%26A...367..876K} {367, 876}

\bibitem[\protect\citeauthoryear{{Kervella}, {Nardetto}, {Bersier}, {Mourard}
  \& {Coud{\'e} du Foresto}}{{Kervella} et~al.}{2004a}]{2004A&A...416..941K}
{Kervella} P.,  {Nardetto} N.,  {Bersier} D.,  {Mourard} D.,   {Coud{\'e} du
  Foresto} V.,  2004a, \mn@doi [\aap] {10.1051/0004-6361:20031743}, \href
  {http://esoads.eso.org/abs/2004A%26A...416..941K} {416, 941}

\bibitem[\protect\citeauthoryear{{Kervella}, {Bersier}, {Mourard}, {Nardetto}
  \& {Coud{\'e} du Foresto}}{{Kervella} et~al.}{2004b}]{2004A&A...423..327K}
{Kervella} P.,  {Bersier} D.,  {Mourard} D.,  {Nardetto} N.,   {Coud{\'e} du
  Foresto} V.,  2004b, \mn@doi [\aap] {10.1051/0004-6361:20035596}, \href
  {http://esoads.eso.org/abs/2004A%26A...423..327K} {423, 327}

\bibitem[\protect\citeauthoryear{{Kervella}, {Bersier}, {Mourard}, {Nardetto},
  {Fouqu{\'e}}  \& {Coud{\'e} du Foresto}}{{Kervella}
  et~al.}{2004c}]{2004A&A...428..587K}
{Kervella} P.,  {Bersier} D.,  {Mourard} D.,  {Nardetto} N.,  {Fouqu{\'e}} P.,
   {Coud{\'e} du Foresto} V.,  2004c, \mn@doi [\aap]
  {10.1051/0004-6361:20041416}, \href
  {http://esoads.eso.org/abs/2004A%26A...428..587K} {428, 587}

\bibitem[\protect\citeauthoryear{{Kervella}, {Fouqu{\'e}}, {Storm}, {Gieren},
  {Bersier}, {Mourard}, {Nardetto}  \& {du Coud{\'e} Foresto}}{{Kervella}
  et~al.}{2004d}]{2004ApJ...604L.113K}
{Kervella} P.,  {Fouqu{\'e}} P.,  {Storm} J.,  {Gieren} W.~P.,  {Bersier} D.,
  {Mourard} D.,  {Nardetto} N.,   {du Coud{\'e} Foresto} V.,  2004d, \mn@doi
  [\apjl] {10.1086/383571}, \href
  {http://esoads.eso.org/abs/2004ApJ...604L.113K} {604, L113}

\bibitem[\protect\citeauthoryear{{Kervella}, {Coud{\'e} du Foresto},
  {Segransan}  \& {di Folco}}{{Kervella} et~al.}{2004e}]{2004SPIE.5491..741K}
{Kervella} P.,  {Coud{\'e} du Foresto} V.,  {Segransan} D.,   {di Folco} E.,
  2004e, in {Traub} W.~A.,  ed.,  Society of Photo-Optical Instrumentation
  Engineers (SPIE) Conference Series Vol. 5491, New Frontiers in Stellar
  Interferometry. p.~741

\bibitem[\protect\citeauthoryear{{Kervella}, {M{\'e}rand}  \&
  {Gallenne}}{{Kervella} et~al.}{2009}]{2009A&A...498..425K}
{Kervella} P.,  {M{\'e}rand} A.,   {Gallenne} A.,  2009, \mn@doi [\aap]
  {10.1051/0004-6361/200811307}, \href
  {http://esoads.eso.org/abs/2009A%26A...498..425K} {498, 425}

\bibitem[\protect\citeauthoryear{{Koen} \& {Eyer}}{{Koen} \&
  {Eyer}}{2002}]{2002MNRAS.331...45K}
{Koen} C.,  {Eyer} L.,  2002, \mn@doi [\mnras]
  {10.1046/j.1365-8711.2002.05150.x}, \href
  {http://adsabs.harvard.edu/abs/2002MNRAS.331...45K} {331, 45}

\bibitem[\protect\citeauthoryear{{Lafrasse}, {Mella}, {Bonneau}, {Duvert},
  {Delfosse}, {Chesneau}  \& {Chelli}}{{Lafrasse}
  et~al.}{2010}]{2010SPIE.7734E..4EL}
{Lafrasse} S.,  {Mella} G.,  {Bonneau} D.,  {Duvert} G.,  {Delfosse} X.,
  {Chesneau} O.,   {Chelli} A.,  2010, in Society of Photo-Optical
  Instrumentation Engineers (SPIE) Conference Series. p.~4 (\mn@eprint {arXiv}
  {1009.0137}), \mn@doi{10.1117/12.857024}

\bibitem[\protect\citeauthoryear{{Le Bouquin} et~al.,}{{Le Bouquin}
  et~al.}{2011}]{2011A&A...535A..67L}
{Le Bouquin} J.-B.,  et~al., 2011, \mn@doi [\aap]
  {10.1051/0004-6361/201117586}, \href
  {http://adsabs.harvard.edu/abs/2011A%26A...535A..67L} {535, A67}

\bibitem[\protect\citeauthoryear{{Leavitt}}{{Leavitt}}{1908}]{1908AnHar..60...87L}
{Leavitt} H.~S.,  1908, Annals of Harvard College Observatory, \href
  {http://esoads.eso.org/abs/1908AnHar..60...87L} {60, 87}

\bibitem[\protect\citeauthoryear{{Leavitt} \& {Pickering}}{{Leavitt} \&
  {Pickering}}{1912}]{1912HarCi.173....1L}
{Leavitt} H.~S.,  {Pickering} E.~C.,  1912, Harvard College Observatory
  Circular, \href {http://adsabs.harvard.edu/abs/1912HarCi.173....1L} {173, 1}

\bibitem[\protect\citeauthoryear{{Lindegren} \& {Dravins}}{{Lindegren} \&
  {Dravins}}{2003}]{2003A&A...401.1185L}
{Lindegren} L.,  {Dravins} D.,  2003, \mn@doi [\aap]
  {10.1051/0004-6361:20030181}, \href
  {http://adsabs.harvard.edu/abs/2003A%26A...401.1185L} {401, 1185}

\bibitem[\protect\citeauthoryear{{Lindemann}}{{Lindemann}}{1918}]{1918MNRAS..78..639L}
{Lindemann} F.~A.,  1918, \mnras, \href
  {http://adsabs.harvard.edu/abs/1918MNRAS..78..639L} {78, 639}

\bibitem[\protect\citeauthoryear{{Luck}, {Andrievsky}, {Kovtyukh}, {Gieren}  \&
  {Graczyk}}{{Luck} et~al.}{2011}]{2011AJ....142...51L}
{Luck} R.~E.,  {Andrievsky} S.~M.,  {Kovtyukh} V.~V.,  {Gieren} W.,   {Graczyk}
  D.,  2011, \mn@doi [\aj] {10.1088/0004-6256/142/2/51}, \href
  {http://adsabs.harvard.edu/abs/2011AJ....142...51L} {142, 51}

\bibitem[\protect\citeauthoryear{{Macri}, {Stanek}, {Bersier}, {Greenhill}  \&
  {Reid}}{{Macri} et~al.}{2006}]{2006ApJ...652.1133M}
{Macri} L.~M.,  {Stanek} K.~Z.,  {Bersier} D.,  {Greenhill} L.~J.,   {Reid}
  M.~J.,  2006, \mn@doi [\apj] {10.1086/508530}, \href
  {http://adsabs.harvard.edu/abs/2006ApJ...652.1133M} {652, 1133}

\bibitem[\protect\citeauthoryear{{Macri}, {Ngeow}, {Kanbur}, {Mahzooni}  \&
  {Smitka}}{{Macri} et~al.}{2015}]{2015AJ....149..117M}
{Macri} L.~M.,  {Ngeow} C.-C.,  {Kanbur} S.~M.,  {Mahzooni} S.,   {Smitka}
  M.~T.,  2015, \mn@doi [\aj] {10.1088/0004-6256/149/4/117}, \href
  {http://adsabs.harvard.edu/abs/2015AJ....149..117M} {149, 117}

\bibitem[\protect\citeauthoryear{{M{\'e}rand}, {Bord{\'e}}  \& {Coud{\'e} du
  Foresto}}{{M{\'e}rand} et~al.}{2005}]{2005A&A...433.1155M}
{M{\'e}rand} A.,  {Bord{\'e}} P.,   {Coud{\'e} du Foresto} V.,  2005, \mn@doi
  [\aap] {10.1051/0004-6361:20041323}, \href
  {http://adsabs.harvard.edu/abs/2005A%26A...433.1155M} {433, 1155}

\bibitem[\protect\citeauthoryear{{M{\'e}rand} et~al.,}{{M{\'e}rand}
  et~al.}{2014}]{2014SPIE.9146E..0JM}
{M{\'e}rand} A.,  et~al., 2014, in Society of Photo-Optical Instrumentation
  Engineers (SPIE) Conference Series. p.~0 (\mn@eprint {arXiv} {1407.2785}),
  \mn@doi{10.1117/12.2057150}

\bibitem[\protect\citeauthoryear{{Nardetto}, {Fokin}, {Mourard}, {Mathias},
  {Kervella}  \& {Bersier}}{{Nardetto} et~al.}{2004}]{2004A&A...428..131N}
{Nardetto} N.,  {Fokin} A.,  {Mourard} D.,  {Mathias} P.,  {Kervella} P.,
  {Bersier} D.,  2004, \mn@doi [\aap] {10.1051/0004-6361:20041419}, \href
  {http://adsabs.harvard.edu/abs/2004A%26A...428..131N} {428, 131}

\bibitem[\protect\citeauthoryear{{Nardetto}, {Mourard}, {Mathias}, {Fokin}  \&
  {Gillet}}{{Nardetto} et~al.}{2007}]{2007A&A...471..661N}
{Nardetto} N.,  {Mourard} D.,  {Mathias} P.,  {Fokin} A.,   {Gillet} D.,  2007,
  \mn@doi [\aap] {10.1051/0004-6361:20066853}, \href
  {http://adsabs.harvard.edu/abs/2007A%26A...471..661N} {471, 661}

\bibitem[\protect\citeauthoryear{{Neilson} \& {Lester}}{{Neilson} \&
  {Lester}}{2013}]{2013A&A...554A..98N}
{Neilson} H.~R.,  {Lester} J.~B.,  2013, \mn@doi [\aap]
  {10.1051/0004-6361/201321502}, \href
  {http://adsabs.harvard.edu/abs/2013A%26A...554A..98N} {554, A98}

\bibitem[\protect\citeauthoryear{{Nordgren}, {Armstrong}, {Germain},
  {Hindsley}, {Hajian}, {Sudol}  \& {Hummel}}{{Nordgren}
  et~al.}{2000}]{2000ApJ...543..972N}
{Nordgren} T.~E.,  {Armstrong} J.~T.,  {Germain} M.~E.,  {Hindsley} R.~B.,
  {Hajian} A.~R.,  {Sudol} J.~J.,   {Hummel} C.~A.,  2000, \mn@doi [\apj]
  {10.1086/317144}, \href {http://esoads.eso.org/abs/2000ApJ...543..972N} {543,
  972}

\bibitem[\protect\citeauthoryear{{Pepe}, {Bouchy}, {Queloz}  \& {Mayor}}{{Pepe}
  et~al.}{2003}]{2003ASPC..294...39P}
{Pepe} F.,  {Bouchy} F.,  {Queloz} D.,   {Mayor} M.,  2003, in {Deming} D.,
  {Seager} S.,  eds,  Astronomical Society of the Pacific Conference Series
  Vol. 294, Scientific Frontiers in Research on Extrasolar Planets. pp 39--42

\bibitem[\protect\citeauthoryear{{Planck Collaboration} et~al.,}{{Planck
  Collaboration} et~al.}{2014}]{2014A&A...571A..16P}
{Planck Collaboration} et~al., 2014, \mn@doi [\aap]
  {10.1051/0004-6361/201321591}, \href
  {http://adsabs.harvard.edu/abs/2014A%26A...571A..16P} {571, A16}

\bibitem[\protect\citeauthoryear{{Poretti}, {Le Borgne}, {Rainer}, {Baglin},
  {Benko}, {Debosscher}  \& {Weiss}}{{Poretti}
  et~al.}{2015}]{2015arXiv150807639P}
{Poretti} E.,  {Le Borgne} J.-F.,  {Rainer} M.,  {Baglin} A.,  {Benko} J.,
  {Debosscher} J.,   {Weiss} W.~W.,  2015, preprint, \href
  {http://adsabs.harvard.edu/abs/2015arXiv150807639P} {} (\mn@eprint {arXiv}
  {1508.07639})

\bibitem[\protect\citeauthoryear{{Queloz} et~al.,}{{Queloz}
  et~al.}{2001}]{2001Msngr.105....1Q}
{Queloz} D.,  et~al., 2001, The Messenger, \href
  {http://adsabs.harvard.edu/abs/2001Msngr.105....1Q} {105, 1}

\bibitem[\protect\citeauthoryear{{Riess} et~al.,}{{Riess}
  et~al.}{2011}]{2011ApJ...730..119R}
{Riess} A.~G.,  et~al., 2011, \mn@doi [\apj] {10.1088/0004-637X/730/2/119},
  \href {http://esoads.eso.org/abs/2011ApJ...730..119R} {730, 119}

\bibitem[\protect\citeauthoryear{{Riess}, {Casertano}, {Anderson}, {MacKenty}
  \& {Filippenko}}{{Riess} et~al.}{2014}]{2014ApJ...785..161R}
{Riess} A.~G.,  {Casertano} S.,  {Anderson} J.,  {MacKenty} J.,   {Filippenko}
  A.~V.,  2014, \mn@doi [\apj] {10.1088/0004-637X/785/2/161}, \href
  {http://adsabs.harvard.edu/abs/2014ApJ...785..161R} {785, 161}

\bibitem[\protect\citeauthoryear{{Sabbey}, {Sasselov}, {Fieldus}, {Lester},
  {Venn}  \& {Butler}}{{Sabbey} et~al.}{1995}]{1995ApJ...446..250S}
{Sabbey} C.~N.,  {Sasselov} D.~D.,  {Fieldus} M.~S.,  {Lester} J.~B.,  {Venn}
  K.~A.,   {Butler} R.~P.,  1995, \mn@doi [\apj] {10.1086/175783}, \href
  {http://adsabs.harvard.edu/abs/1995ApJ...446..250S} {446, 250}

\bibitem[\protect\citeauthoryear{{Scowcroft}, {Freedman}, {Madore}, {Monson},
  {Persson}, {Rich}, {Seibert}  \& {Rigby}}{{Scowcroft}
  et~al.}{2015}]{2015arXiv150206995S}
{Scowcroft} V.,  {Freedman} W.~L.,  {Madore} B.~F.,  {Monson} A.,  {Persson}
  S.~E.,  {Rich} J.,  {Seibert} M.,   {Rigby} J.~R.,  2015, preprint, \href
  {http://adsabs.harvard.edu/abs/2015arXiv150206995S} {} (\mn@eprint {arXiv}
  {1502.06995})

\bibitem[\protect\citeauthoryear{{Soszynski} et~al.,}{{Soszynski}
  et~al.}{2008}]{2008AcA....58..163S}
{Soszynski} I.,  et~al., 2008, \actaa, \href
  {http://adsabs.harvard.edu/abs/2008AcA....58..163S} {58, 163}

\bibitem[\protect\citeauthoryear{{Storm}, {Carney}, {Gieren}, {Fouqu{\'e}},
  {Latham}  \& {Fry}}{{Storm} et~al.}{2004}]{2004A&A...415..531S}
{Storm} J.,  {Carney} B.~W.,  {Gieren} W.~P.,  {Fouqu{\'e}} P.,  {Latham}
  D.~W.,   {Fry} A.~M.,  2004, \mn@doi [\aap] {10.1051/0004-6361:20034634},
  \href {http://adsabs.harvard.edu/abs/2004A%26A...415..531S} {415, 531}

\bibitem[\protect\citeauthoryear{{Storm} et~al.,}{{Storm}
  et~al.}{2011}]{2011A&A...534A..94S}
{Storm} J.,  et~al., 2011, \mn@doi [\aap] {10.1051/0004-6361/201117155}, \href
  {http://adsabs.harvard.edu/abs/2011A%26A...534A..94S} {534, A94}

\bibitem[\protect\citeauthoryear{{Stoy}}{{Stoy}}{1959}]{1959MNSSA..18...48S}
{Stoy} R.~H.,  1959, Monthly Notes of the Astronomical Society of South Africa,
  \href {http://adsabs.harvard.edu/abs/1959MNSSA..18...48S} {18, 48}

\bibitem[\protect\citeauthoryear{{Suyu} et~al.,}{{Suyu}
  et~al.}{2012}]{2012arXiv1202.4459S}
{Suyu} S.~H.,  et~al., 2012, preprint, \href
  {http://adsabs.harvard.edu/abs/2012arXiv1202.4459S} {} (\mn@eprint {arXiv}
  {1202.4459})

\bibitem[\protect\citeauthoryear{{Tallon-Bosc} et~al.,}{{Tallon-Bosc}
  et~al.}{2008}]{2008SPIE.7013E..1JT}
{Tallon-Bosc} I.,  et~al., 2008, in Society of Photo-Optical Instrumentation
  Engineers (SPIE) Conference Series. p.~1, \mn@doi{10.1117/12.788871}

\bibitem[\protect\citeauthoryear{{Wesselink}}{{Wesselink}}{1946}]{1946BAN....10...91W}
{Wesselink} A.~J.,  1946, \bain, \href
  {http://adsabs.harvard.edu/abs/1946BAN....10...91W} {10, 91}

\bibitem[\protect\citeauthoryear{{van Leeuwen}, {Feast}, {Whitelock}  \&
  {Laney}}{{van Leeuwen} et~al.}{2007}]{2007MNRAS.379..723V}
{van Leeuwen} F.,  {Feast} M.~W.,  {Whitelock} P.~A.,   {Laney} C.~D.,  2007,
  \mn@doi [\mnras] {10.1111/j.1365-2966.2007.11972.x}, \href
  {http://adsabs.harvard.edu/abs/2007MNRAS.379..723V} {379, 723}

\makeatother
\end{thebibliography}

\appendix 
\section{Supporting Evidence for Modulated Angular
Variability}\label{sec:obs:pionier:standards}\label{app:A} This appendix
provides details of the tests we carried out to investigate the precision of the
diameters inferred from our interferometric measurements and the instrumental
stability. This is a crucial step in determining whether the minute
cycle-to-cycle differences between the maximal diameters subtended by \lc\ are
real or explained by systematics. While the accuracy of angular diameter
measurements is dominated by the accuracy of the wavelength calibration and the
treatment of limb darkening, significantly higher {\it precision} may be
achieved by eliminating sources of bias via a differential measurement. The
tentative evidence for such cycle-to-cycle variations presented here relies on
such a differential measurement involving the mean maximal diameters determined
during the consecutive runs A and C.

The following elements involved in the observation affect the precision of this
differential measurement. These can be separated into effects acting on
intra-run time-scales ($< 10$ d) and inter-run time-scales ($> 10$ d).
Intra-run effects include
\begin{enumerate}
  \item nightly wavelength calibrations (can be traced via standard stars);
  \item choice of calibrator stars (avoided by using common set of calibrator
  stars);
  \item sampling of UV-plane, i.e., projected baselines;
  \item telescope or instrument vibrations (should mainly cancel out over
  half-nights of observations);
  \item pupil stability (affects baseline stability).
\end{enumerate}

Any biases introduced by such effects should affect all stars observed during a
given night in the same way. Standard star observations, i.e., treating
calibrator stars of known high stability as science targets, thus provide a
means to monitor the stability and the precision that can be achieved during a
given observing run. If the same standard star was observed during multiple
observing runs, this information can be used to track instrumental stability
over a longer timeframe. We perform such a test in
Appendix\,\ref{app:sec:standards} below, finding excellent inter-run precision for standard star HD\,74088 and no
significant offsets between runs B and C for standard star HD\,81101.

Inter-run effects (mainly introduced by observations at different azimuthal
angles) include the following.
\begin{enumerate}
 \item All intra-run effects listed above.
  \item Stellar companions to \lc\, although Sect.\,\ref{sec:disc:companion}
  excludes this possibility.
  \item Stellar companions to calibrator stars, in analogy with the effect
  stated above for the science target. Using more than one calibrator per run
  (as we did) reduces the impact of such an effect. We additionally inspect the
  closure phase measurements for all standard stars in Appendix\,\ref{app:sec:cp} below. 
  \item Stellar ablation due to rotation (calibrators and science target) can
 yield similar intra-run differences due to different viewing angles. However,
 rotation is expected to be very slow for all calibrator stars used
 (late-type giant stars) as well as for \lc . Nevertheless, \lc 's $v\sin{i}
 \approx 7$\,\kms\ may lead to a flattening on the order of $0.2-0.3\%$
 (assuming a Roche model with $8$\,\Msol\ and $180$\,\Rsol), i.e., around
 $7.5\,\mu$as. 
 \item Possible surface inhomogeneities (spots).
 \item Changes in the instrumental setup: mirror coatings (ageing effects,
 re-coatings of individual mirrors), optical defects, polarization. 
 \item asymmetric circumstellar envelopes (CSE). CSEs mainly affect short
 baseline and the top part of the $V^2$ curve. We therefore discarded short
 baselines and re-determined diameters with only long baselines, but found
 virtually no difference with the result based on all baselines, cf.
 Appendix\,\ref{app:sec:shortremoved}.
 \item Different baseline configurations used. One AT was located at different
 stations during runs A (station I1) and C (station J3). We investigate whether
 this result might introduce a bias in Appendix\,\ref{app:sec:I1J3}, and find no
 sign of a bias.
\end{enumerate}

Unfortunately, no standard star observations are available for run A, precluding
a final assessment of any inter-run differences between run A and run C. While
all of our tests indicate that there is no cause for alarm\footnote{This one is
for VLT observers}, we acknowledge that there remains room for systematics that
can lead to a spurious result, since we cannot \emph{demonstrate} the absence of
inter-run biases between run A and C. We thus conservatively consider the
evidence for modulated angular diameter variability as tentative and call for
prudence in the interpretation of this result. To improve this situation in
the future, we plan to observe the standard stars HD\,74088, HD\,81011 during
all future observing runs in order to identify and monitor instrumental effects
and distinguish them clearly from real modulated variability.

\subsection{Intra- and Inter-run Stability of Inferred Diameters determined
using Standard stars} 
\label{app:sec:standards}

\begin{figure}
\centering
\includegraphics[scale=0.95]{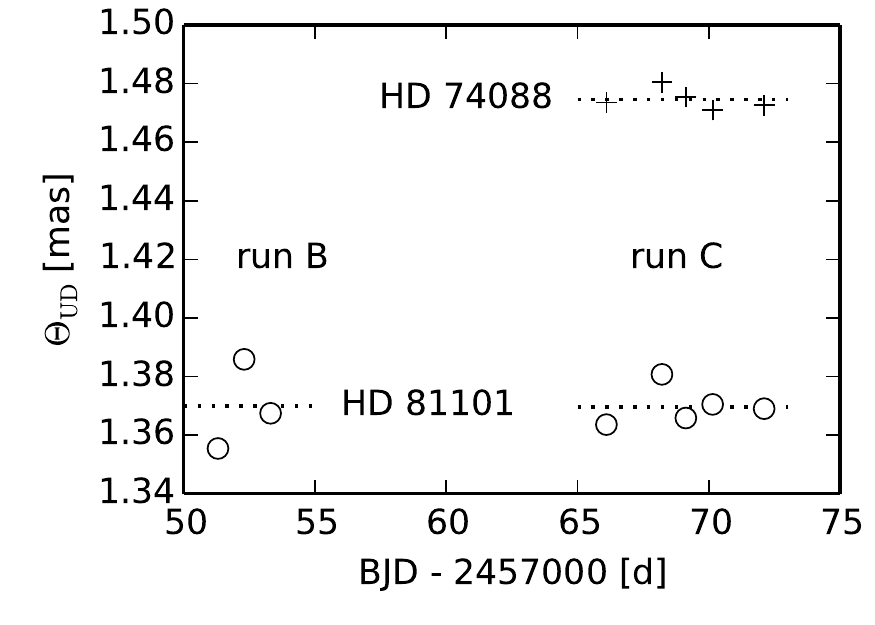}
\caption{Time series of angular diameters for standard stars. HD\,74088 is
plotted as black pluses and indicates excellent intra-run stability ($\sim 0.0015$\, mas). HD\,81101 is plotted as black open circles and can be used to
trace inter-run stability. There is no significant difference between runs B and C for HD\,81101 and HD\,74088 and this demonstrates excellent intra-run stability.}
\label{app:fig:VLTIstandards}
\end{figure}

Using the calibrator star HD\,74088 as a \emph{standard}\footnote{i.e., we
treated it as a science object, calibrating the visibilities with the same
stars that were used for \lc.} star, we investigate the stability of UD diameters
over the timespan of one run, specifically run C. Figure\,\ref{app:fig:VLTIstandards} shows the UD diameters determined with {\tt LITPro} as black pluses. We find $\langle \Theta \rangle = 1.4746 \pm 0.0015$\, mas over the five nights. This indicates excellent intra-run stability and corroborates the use of mean diameters for investigating the modulated variability of \lc . 

Analogously, we use HD\,81101 (observed during runs B and C) to test the
inter-run stability of UD diameters. Figure\,\ref{app:fig:VLTIstandards} shows
these diameters as open circles. We find no significant difference between the
diameters determined during runs B and C, with a difference of $0.0004 \pm
0.0068$. The absence of any inter-run differences corroborates  the
apparent cycle-to-cycle difference of the maximal diameters.

Unfortunately, we do not have a sufficient number of calibrator stars available
to perform this test for runs A and C. We did, however, perform a test where we
calibrated each calibrator with the other and checked for the stability of the
result. Of course, the results of this test are not independent, since a bias in
either calibrator will directly affect the `science' target. Nevertheless, we
find no clear signs of systematic differences between runs A and C; the offsets
between both runs are each approximately $8\,\mu$as, with opposite sign. We find
standard mean errors during each run of $3-5\,\mu$as for both stars.
This example shows the importance of observing additional calibrator stars as
standards to trace the instrumental stability.

\subsection{Removing short baselines with $V^2 > 0.5$ for \lc}
\label{app:sec:shortremoved} 
\begin{figure}
\centering
\includegraphics[scale=0.95]{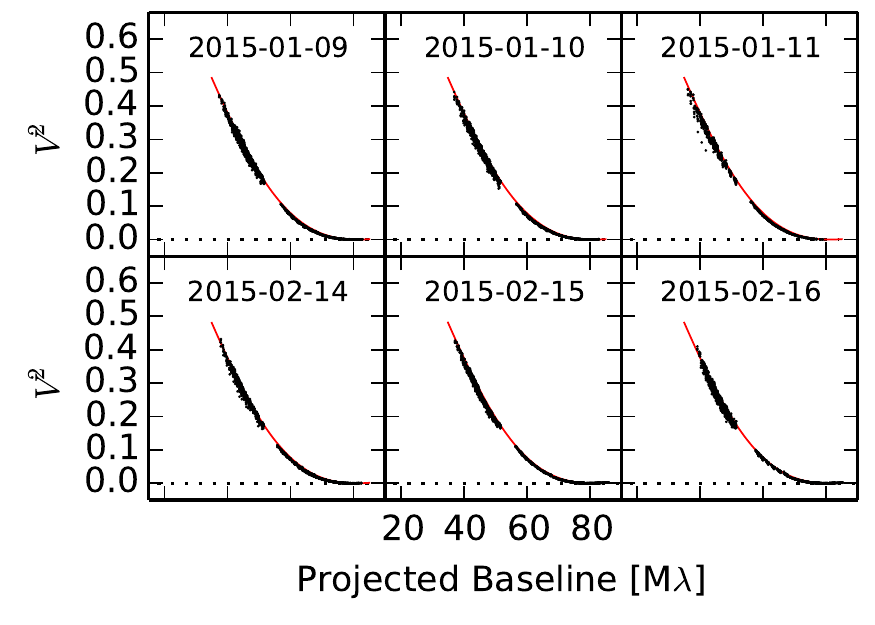}
\caption{Visibility curves with short baselines removed for nights near maximum
diameter.}
\label{app:fig:visshortremoved}
\end{figure}

Here we test the sensitivity to a possible circumstellar environment by
discarding measurements at short baselines. We list the inferred UD diameters in 
Tab.\,\ref{app:tab:visshortremoved} and show the $V^2$ curves in
Fig.\,\ref{app:fig:visshortremoved}. We obtain results that are virtually
identical to the UD diameters in Tab.\,\ref{tab:vlti:diams} and find a clear
(formal) difference between the diameters at the two maxima.

\begin{table}
\begin{tabular}{@{}l|rrr|rrr@{}}
\hline
Night & 01-09 & 01-10 & 01-11 & 02-14 & 02-15 & 02-16 \\
\hline
$\Theta_{\rm{UD}}$\ [mas] & 3.0979 & 3.0974 & 3.0970 & 3.1118 & 3.1089 &
3.1105 \\
$\sigma(\Theta_{\rm{UD}})$\ [mas] & 0.0003 & 0.0003 & 0.0003 & 0.0003 & 0.0002
& 0.0003 \\
$\chi^2_r$ & 2.533 & 1.763 & 1.847 & 3.363 & 3.204 & 3.896 \\
\hline
$\langle \Theta_{\rm{UD}} \rangle$\ [mas] & \multicolumn{3}{c}{$3.0974 \pm
0.0002$} & \multicolumn{3}{c}{$3.1104 \pm 0.0006$} \\
$\Delta \langle \Theta_{\rm{UD}} \rangle$\ [mas] & \multicolumn{6}{c}{$0.013\pm
0.001$ } \\
\hline
\end{tabular}
\caption{Uniform disk diameters determined for nights near maximum with short
baseline removed. The three-night averages are nearly identical to the results
in Tab.\,\ref{tab:vlti:meandiams} and indicate a larger second maximum.}
\label{app:tab:visshortremoved}
\end{table}

\subsection{Discarding stations I1 and J3 for \lc}\label{app:sec:I1J3}

\begin{figure}
\centering
\includegraphics[scale=0.95]{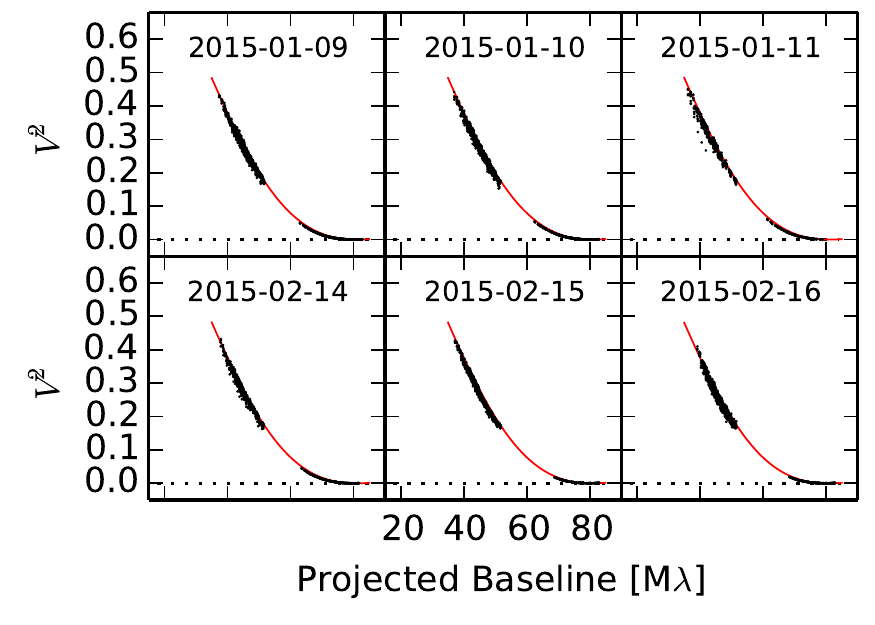}
\caption{Visibility curves with baselines I1 (run A) and J3 (run C) removed for
nights near maximum diameter.}
\label{app:fig:visI1J3removed}
\end{figure}

One of the ATs was positioned at station I1 during run A and at station J3
during run C. Here we remove all baselines involving that telescope to test
whether this difference could bias our result. We show the resulting $V^2$
curves in Fig.\,\ref{app:fig:visI1J3removed} and list the UD diameters in
Tab.\,\ref{app:tab:I1J3}. The resulting diameters are in agreement with the
results including all baselines, and the difference between the maximal
diameters remains clear.

\begin{table}
\begin{tabular}{@{}l|rrr|rrr@{}}
\hline
Night & 01-09 & 01-10 & 01-11 & 02-14 & 02-15 & 02-16 \\
\hline
$\Theta_{\rm{UD}}$\ [mas] &  3.0999 & 3.0978 & 3.0976 & 3.1066 & 3.1080 &
3.1106\\
$\sigma(\Theta_{\rm{UD}})$\ [mas] & 0.0004 & 0.0003 & 0.0004 & 0.0003 &
0.0002 & 0.0003 \\
$\chi^2_r$ &  2.603 & 1.868 & 2.152 & 1.853 & 2.623 & 3.140 \\
\hline
$\langle \Theta_{\rm{UD}} \rangle$\ [mas] & \multicolumn{3}{c}{$3.0985 \pm
0.0006$} & \multicolumn{3}{c}{$3.1084 \pm 0.0010$} \\
$\Delta \langle \Theta_{\rm{UD}} \rangle$\ [mas] & \multicolumn{6}{c}{$0.010\pm 0.001$ } \\
\hline
\end{tabular}
\caption{Uniform disk diameters determined for nights near maximum with all I1
(run A) and J3 (run C) baselines removed. The three-night averages show a 
clear difference, indicating a larger second maximum.}
\label{app:tab:I1J3}
\end{table}

\subsection{Closure phase stability}
\label{app:sec:cp}

Figures\,\ref{app:fig:t3phiHD81502} through \ref{app:fig:t3phiHD81101} present the stability of closure phases for all calibrator stars. To
within the precision of \Pionier , none of the calibrator/standard
stars shows signs of asymmetry or companions \citep[known
binaries were rejected by][and none of the calibrators
are listed as multiples in \citealt{2005A&A...433.1155M}]{2010SPIE.7734E..4EL}.
The absence of time-dependent changes suggests that the changing azimuthal angle of the observations does not
introduce a bias for the diameter determination.
\begin{figure}
\centering
\includegraphics[scale=0.95]{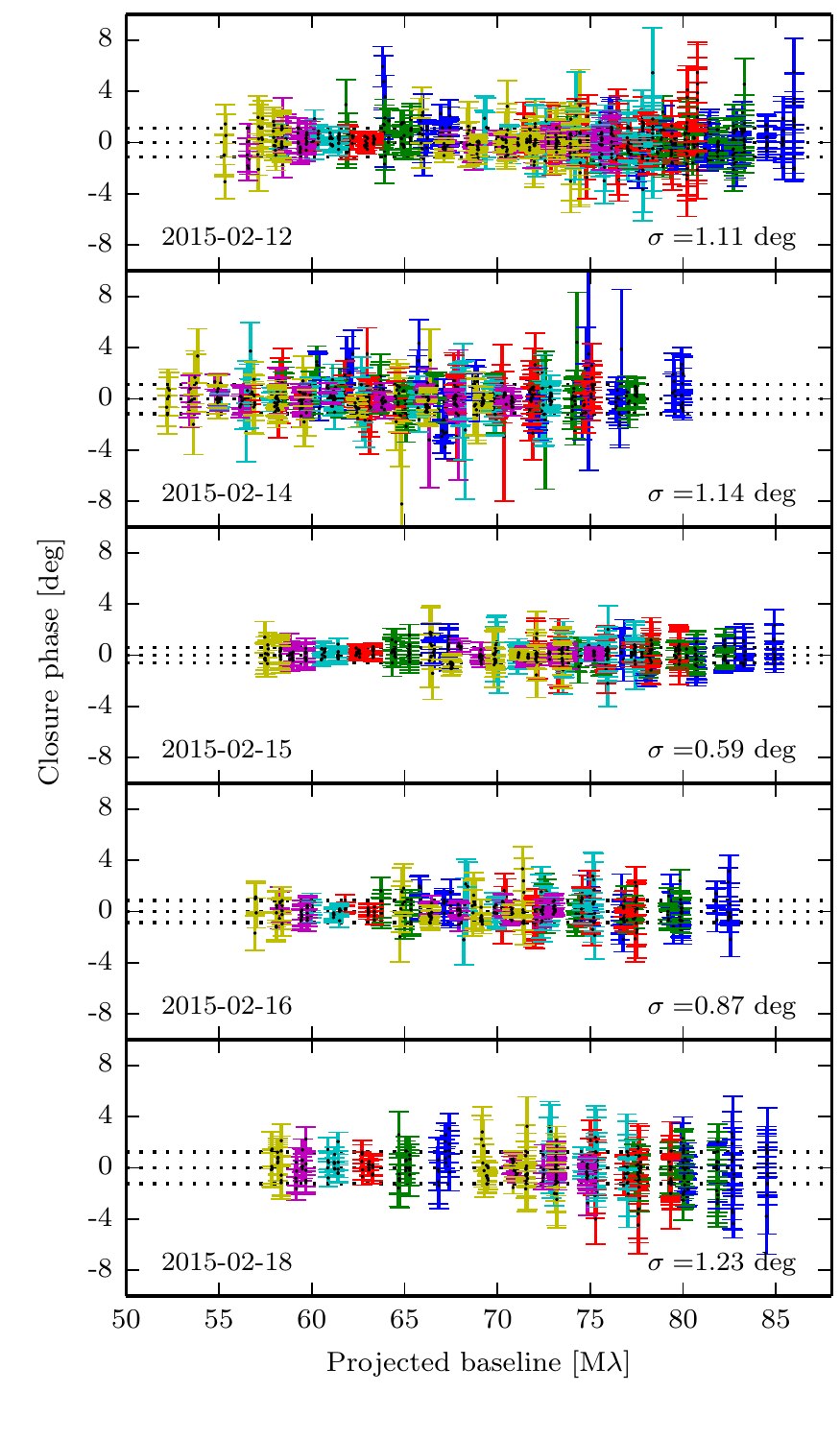}
\caption{Closure phases versus projected baseline for standard star HD\,74088. All
panels have identical axis ranges. The dotted lines indicate 0 degrees and the
$1\sigma$ range, which is also given in the bottom right of each panel. The
night of observation is printed in the bottom left of each panel. Different colors represent different spectral channels.}
\label{app:fig:t3phiHD74088}
\end{figure}

\begin{figure*}
\centering
\includegraphics{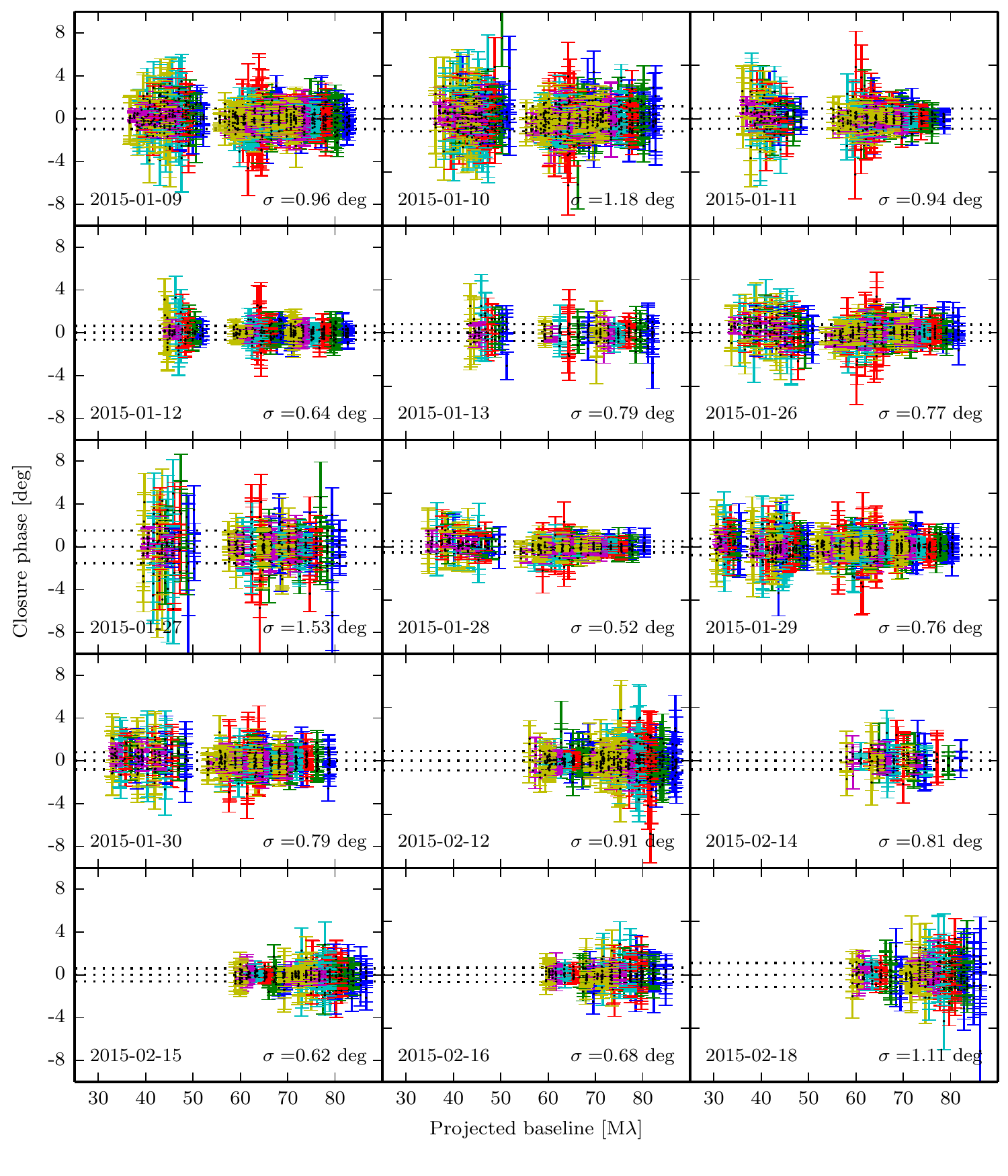}
\caption{Closure phases versus projected baseline for calibrator star HD\,81502, cf.
also Fig.\ref{app:fig:t3phiHD74088}.}
\label{app:fig:t3phiHD81502}
\end{figure*}

\begin{figure*}
\centering
\includegraphics{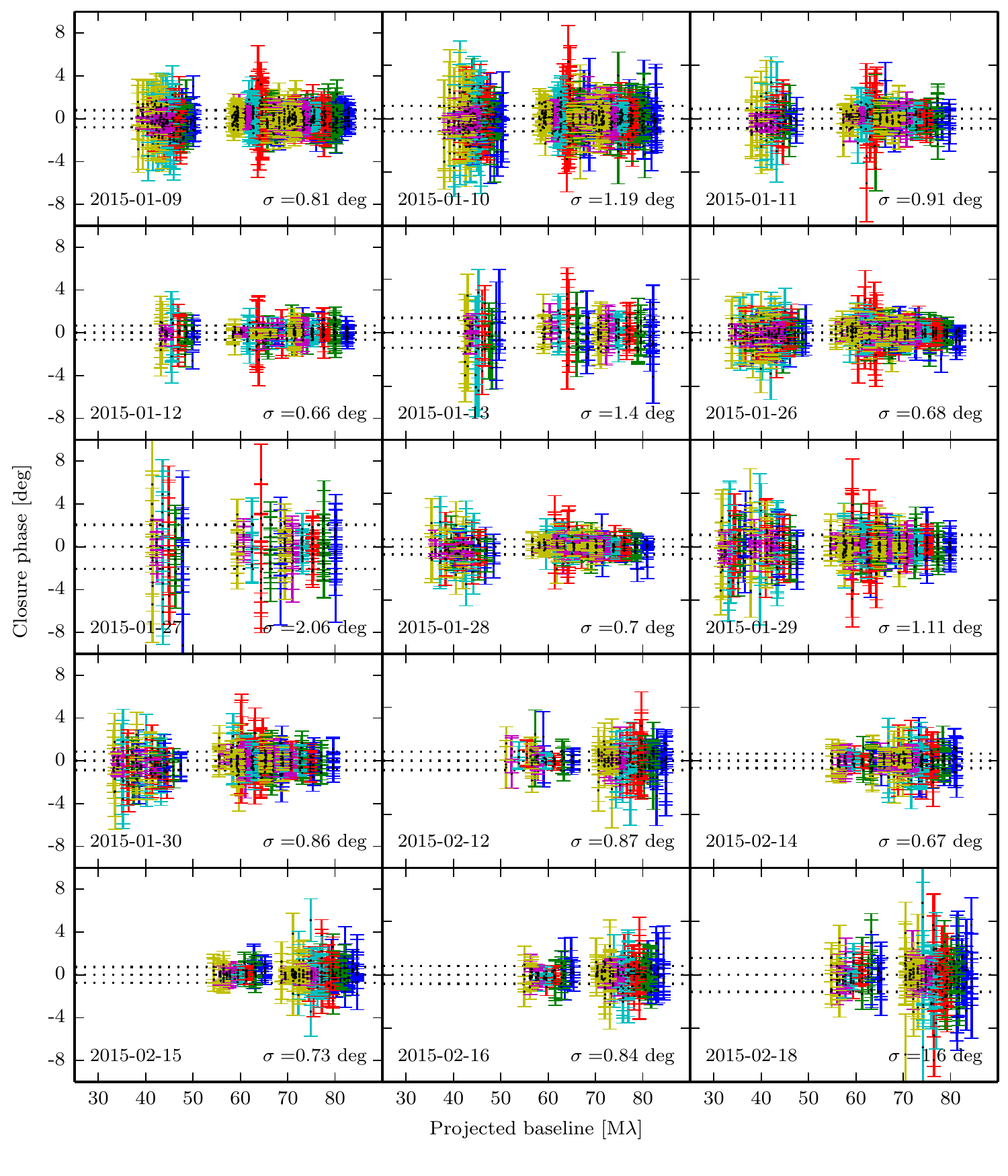}
\caption{Closure phases versus projected baseline for calibrator star HD\,89805, cf.
also Fig.\ref{app:fig:t3phiHD74088}.}
\label{app:fig:t3phiHD89805}
\end{figure*}

\begin{figure*}
\centering
\includegraphics{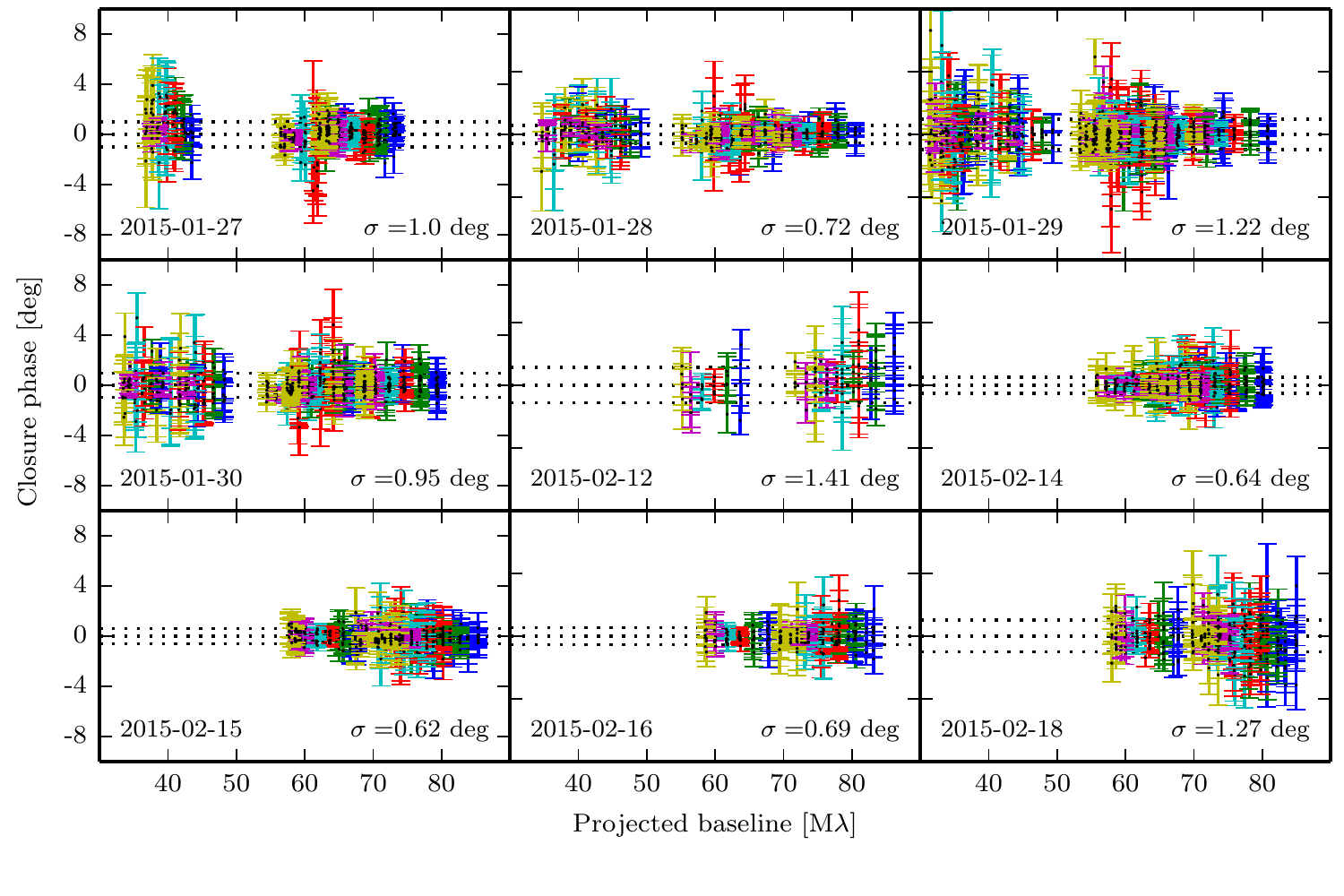}
\caption{Closure phases versus projected baseline for standard star HD\,81101, cf.
also Fig.\ref{app:fig:t3phiHD74088}.}
\label{app:fig:t3phiHD81101}
\end{figure*}

\label{lastpage}

\end{document}